\documentclass[aps,pra,twocolumn,amsmath,amssymb,superscriptaddress,reprint,longbibliography]{revtex4-1}

\usepackage[utf8]{inputenc}
\usepackage{graphicx}
\usepackage[colorlinks=true,citecolor=blue]{hyperref}
\usepackage[caption=false]{subfig}
\newcommand\MLE{\text{\tiny{MLE}}}
\newcommand\CS{\text{\tiny{CS}}}
\newcommand\CC{\text{\tiny{CC}}}
\newcommand\HT{\text{\tiny{HT}}}
\newcommand\GT{\text{\tiny{GT}}}
\newcommand\MS{\mathcal{S}}

\begin{document}
\title{An improved estimator of Shannon entropy with applications to systems with memory}
\author{Juan De Gregorio}
\author{David S\'anchez}
\author{Ra\'ul Toral}
\address{
 Institute for Cross-Disciplinary Physics and Complex Systems IFISC (UIB-CSIC), Campus Universitat de les Illes Balears, E-07122 Palma de Mallorca, Spain
}
\date{\today}
\begin{abstract}
We investigate the memory properties of discrete sequences built upon a finite number of states. We find that the block entropy can reliably determine the memory for systems modeled as Markov chains of arbitrary finite order. Further, we provide an entropy estimator that remarkably gives accurate results when correlations are present. To illustrate our findings, we calculate the memory of daily precipitation series at different locations. Our results are in agreement with existing methods being at the same time valid in the undersampled regime and independent of model selection.
\end{abstract}

\maketitle

\section{\label{sec:level1} Introduction }

Stochastic modeling lies at the heart of many modern studies of complex systems. In this approach, the detailed dynamics is replaced by a set of transition probabilities that connect the states of the system at two different times. Quite often, the resulting stochastic processes can be modeled with the aid of discrete Markov chains~\cite{cox17}, where the transition probabilities to a future state depend only on the present state and not on past states~\cite{markov}. The broad applicability of this Markovian dynamics runs across disciplines: fluctuation theory in statistical physics~\cite{gar04}, DNA sequence analysis in molecular biology~\cite{chu89}, weather forecast in meteorology simulations~\cite{wil99}, or web searches in information retrieval~\cite{pag06}, just to mention a few.

However, this memoryless approximation is insufficient if one deals with strongly correlated systems~\cite{han82,may07}, whose next state depends not only on the present one but also on a longer history of past states. Higher-order Markov chains~\cite{raf85}, which include a finite number $m$ of past states to determine the transition probability to the next one, are thus needed for an accurate characterization of systems with memory. Recent examples of systems modeled with higher-order Markovian discrete chains include Parkinsonian tremor time series~\cite{yul06}, linguistically detected emotional events~\cite{ho12}, genomic polymorphisms for cancer research~\cite{sei12}, human navigation on the web~\cite{sin14} and autochemotactic searchers~\cite{mey21}. It is therefore of utmost importance to have reliable methods that faithfully determine the order or memory $m$ for a given sequence of data \cite{crutchfield}.

This goal can be achieved in several ways. For the purpose of this paper, we focus on information-theoretical methods. The widely used Akaike information criterion (AIC)~\cite{aka98} is based on a loss function constructed from the Kullback-Leibler relative entropy. The loss function is then minimized to find a good estimate of $m$~\cite{ton75}. Since this approach views the transition probabilities from a frequentist perspective, the Bayesian information criterion (BIC) uses instead Bayes factors to further constrain the loss function~\cite{kat81}. While the AIC criterion is shown to overestimate $m$, the BIC is biased towards overly simple models~\cite{wea99}. To overcome these difficulties we propose in this paper a new method to determine the order of a Markov chain that best represents a given series of data. Advantages of the proposed method include that it is independent on model selection and sufficiently precise even in the undersampled regime.

Central to our proposal is the block Shannon entropy~\cite{sch97}. Data are grouped in blocks of a given size $n$ from which the entropy is extracted using the probability of each block. When the total sequence length is much larger than the number of possible blocks, such that all blocks appear with sufficient statistical relevance, the maximum likelihood estimator for the entropy works well. Nevertheless, sequences extracted from real data can be short and the previous estimator is consequently negatively biased~\cite{chao}. There have been several attempts to derive unbiased estimators for the entropy but, to the best of our knowledge, all these proposals rely on the sequence being composed by independent random states. What is needed is a faithful estimator being able to deal with correlated systems, as inferred in this work.

Therefore, our contribution is twofold. First, we suggest a novel method that determines the memory of a generic discrete sequence when the series is described by a Markov chain of arbitrary order. Second, we introduce an estimator suitable for the expected value of any observable (like the entropy) that provides information of a correlated system. We emphasize that both findings are independent and constitute, by their own, interesting advances that do not rely on each other. The calculation technique for the order of Markov chains would still hold if another block entropy estimator is used in any of the multiple contexts where these processes are relevant~\cite{hag02}. Additionally, our improved estimator can be applied to any model of correlations (not necessarily a Markov chain) and is thus of interest in situations where entropy is fundamental to understand the system's properties~\cite{pom94}.

The paper is organized as follows. In Section~\ref{sec:method} we develop the main relation between the block entropy and the order of a Markov chain. We then explain the method to extract the order from the entropy estimator, based on a linear relation between the block entropy and the order, whose proof is left for the Appendix. In Section~\ref{sec:entropy} we develop the algorithm to estimate the block entropy from a series of correlated data. In Section~\ref{sec:results-numerical} we present a case study where we determine the memory of a series constructed numerically, while in Section~\ref{sec:results-daily} we apply the method to determine the order of daily precipitation series, comparing with the results found within the BIC approach. Finally, Section~\ref{sec:conclusions} contains our conclusions. 

\section{Relation between block entropy and memory}
\label{sec:method}

\begin{figure}[t]
\includegraphics[width=0.9\columnwidth]{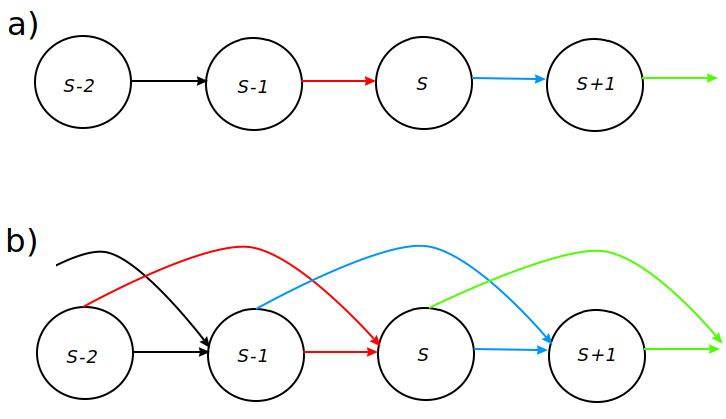}
\caption{Sketch for the transition probabilities of two sequences with memory {a)} $m=1$ and {b)} $m=2$.}
\label{fig:diagram}
\end{figure}

Consider a discrete random variable $X$ with $L$ possible outcomes, $\{z_i\}, i \in\{1,\ldots,L\} $.
Let ${\cal S}=(X_{1},\ldots,X_{N})$ be a sequence of time-ordered observations, where $N$ is the sequence size. Then, ${\cal S}$ has memory $m\ge 1$ if the transition probabilities satisfy
\begin{equation}
\begin{split}
P(X_{s}=z_j|X_{s-1}=z_{l},\ldots, X_{1}=z_{k}) = \\
P(X_{s}=z_{j}|X_{s-1}=z_{l},\ldots,X_{s-m}=z_{r}),
\label{eq:transProbs}
\end{split}
\end{equation}
with $s$ the time or position in the series. For the case $m=0$ the probabilities $P(X_{s}=z_j)$ are independent for all $s$. The sequence is termed Markovian if $m=1$ in Eq.~\eqref{eq:transProbs}.
Values $m>1$ correspond to higher-order Markov systems, which are the focus of this work. 
In Fig.~\ref{fig:diagram}(a) we illustrate the transition probabilities for a generic sequence with $m=1$: the outcome probability at time $s$ depends only on the previous state at time $s-1$ (red arrow).
Figure~\ref{fig:diagram}(b) is a graphical representation for the case $m=2$, where now the probability to observe the system in a given state at time $s$ depends on states at {\em both} previous times $s-1$ and $s-2$ (red arrows).

In what follows, we consider homogeneous Markov chains for which the $L^{m+1}$ transition probabilities of Eq.~\eqref{eq:transProbs} are independent of time index $s$ and the conditional probabilities are then uniquely determined by the states $\{z_i\}$. Furthermore, we will only consider the stationary case and therefore neglect transient effects.

Given a long sequence ${\cal S}$ it is not an easy task to determine the order $m$ by a direct application of Eq.~\eqref{eq:transProbs} since one should check a growing number of relations as $m$ is increased. However, a much more efficient procedure stems from an analysis of the block Shannon entropy \cite{shannon,par06}. Let us form overlapping blocks in ${\cal S}$ of size ${n\ge 1}$. The $j$-th block is then $B_{j}^{(n)} = (X_j,\ldots,X_{j+n-1})$ with $1\leq j \leq N-n+1\equiv N_n$. There exist $L^n$ distinct possible blocks of size $n$, which we denote as $\lbrace b_{i}^{(n)} \rbrace_{1 \leq i \leq L^n}$. All observed blocks $\{B_j^{(n)}\}_{j=1,\dots,N_n}$ in the series $\MS$ belong to the set $\lbrace b_{i}^{(n)} \rbrace_{1 \leq i \leq L^n}$. The block Shannon entropy is defined by
\begin{equation}
H_n = -\sum_{i=1}^{L^n} p(b_{i}^{(n)})\log(p(b_{i}^{(n)})),\quad n\ge 1,
\label{eq:HBSE}
\end{equation}
where $p(b_i^{(n)})$ is the probability of appearance of the block sequence $b_i^{(n)}$. We use the convention that $H_0\equiv 0$.

The key point of the proposed method to determine the order of a stochastic process is to realize that $H_{n}$ is a linear function of $n$ for $n \geq m$ if and only if the sequence has memory $m$ (see \ref{sec:appA} for a proof).
Therefore, we can write
\begin{equation}
\label{Hnlineal}
H_{n} = (H_{m+1}-H_{m})(n-m)+H_{m} \quad n \geq m \ge 0,
\end{equation}
which, for the independent case ($m=0$), reduces to the known expression for the entropy rate per symbol~\cite{cover}
\begin{equation}
H_{n} = nH_{1} \quad n \geq 1.
\end{equation}

For the general case $m>0$ the following procedure yields an accurate assessment of $m$. Given a trial memory $\mu=0,1,2,\dots$ we define the {\slshape trial entropies} $\mathcal{H}_{\mu}(n)$ by the relation:
\begin{equation}
\mathcal{H}_{\mu}(n) = (H_{\mu+1}-H_{\mu})(n-\mu)+H_{\mu} \quad n \geq \mu.
\label{eq:Hmu}
\end{equation}
Now, according to Eq.~\eqref{Hnlineal}, $\mathcal{H}_\mu(n)$ would be equal to $H_n$ for $n\ge \mu$ if the series $\MS$ could be represented by a Markov chain of order $\mu$. 
Therefore, if we consider the mean squared error
\begin{equation}
\Delta_{\mu} = \dfrac{1}{n_{\text{max}}-\mu+1}\sum_{n=\mu}^{n_{\text{max}}}(\mathcal{H}_{\mu}(n)-H_{n})^2,
\label{eq:Deltamu}
\end{equation}
we find that $\Delta_\mu$ vanishes for $\mu \ge m$, independently of the value of the cutoff $n_{\max}$. Hence, the criterion to find the order $m$ of the sequence $\MS$ is
\begin{equation} \label{eq:m}
m=\text{min}(\lbrace\mu:\Delta_{\mu}=0\rbrace).
\end{equation}
In contrast to the BIC and AIC methods, where the loss function depends on the model selected, Eq.~\eqref{eq:m} is independent of any selection procedure. For memory determination, both AIC and BIC methods are generally used along with a comparison among possible models suitable for a given sequence, whereas our proposed method provides an exact result for discrete random variables. If the system is naturally described with a continuous random variable, we first need to discretize this variable and the results obtained with Eq.~\eqref{eq:m} will depend on the binsize used for the discretization. 

\begin{figure}[t]
\includegraphics[width=0.9\columnwidth]{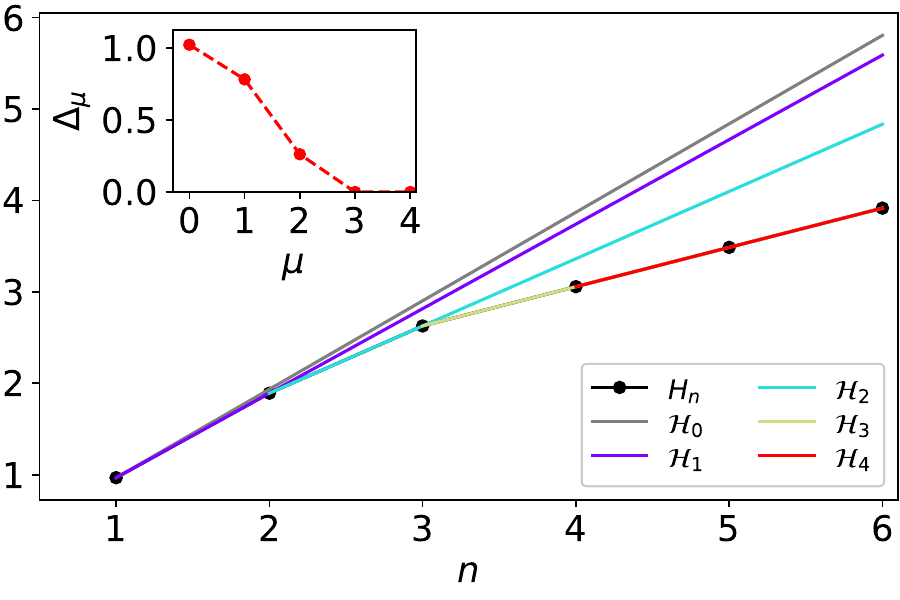}
\caption{Block entropies $H_n$ (dots) and $\mathcal{H}_{\mu}(n)$ ($\mu = 0,\ldots,4$ with different colors) versus block size $n$ for a binary sequence, $L=2$, with memory $m=3$ and transition probabilities chosen randomly from the interval $(0,1)$. Inset: Mean squared error $\Delta_\mu$ as a function of the trial memory $\mu$. The proposed method determines the true memory from the condition $\text{min}(\lbrace\mu:\Delta_{\mu}=0\rbrace)$, which is satisfied at $\mu=3$.}
\label{fig:method}
\end{figure}

As an example of an explicit calculation of $m$, in Fig.~\ref{fig:method} we show results for a binary system ($L=2$) and memory $m=3$. The $2^{3+1}=16$ transition probabilities are chosen randomly from a uniform distribution in the interval $(0,1)$ (and multiplied by a common factor in order to fulfill the necessary normalization conditions). We plot in the main panel with color lines the trial entropy $\mathcal{H}_\mu(n)$ given by Eq.~\eqref{eq:Hmu} for different values of $\mu$. For comparison, the exact block entropies $H_n$, computed from the transition probabilities, are also shown with black dots. Clearly, the trial entropies $\mathcal{H}_\mu(n)$ coincide for $\mu\ge 3$ with the exact block entropy $H_n$ in agreement with the memory of the process. Alternatively, the inset of Fig.~\ref{fig:method} shows that the mean squared error $\Delta_\mu$ vanishes for $\mu\ge 3$. Therefore, Eq.~\eqref{eq:m} is met and the memory is $m=3$, as expected.

This particular example is based on a particular set of transition probabilities drawn from a uniform distribution. We stress that for a different set of transition probabilities the curves in Fig.~\ref{fig:method} would change quantitatively but the condition $\Delta_{\mu} = 0$ for $\mu \geq 3$ will always hold for a system with memory $m=3$. As a consequence, the method is robust against strong variations of the transitions between states.

It should be clear that in this example the determination of the memory $m$ has been very efficient because we had access to the true block entropies $H_n$, determined from the knowledge of the $L^{m+1}$ transition probabilities. In a case where this is not possible, for example, if we are given a numerical sequence $\MS$ of finite length $N$, we would need to replace the exact block entropies $H_n$ in Eqs.~\eqref{eq:Hmu} and \eqref{eq:Deltamu} with appropriate estimators $\hat{H}_n$. As a consequence, a good performance of our method requires a reliable estimator of the block entropies. Unfortunately, this is not an easy issue since all the entropy estimators reported in the literature have systematic biases that degrade the performance of the method. In the next section, we will propose a new entropy estimator that turns out to be particularly suitable for our purposes of determining the memory of a higher-order Markov process.

\section{Entropy estimator}
\label{sec:entropy}
Block probabilities and, hence, the Shannon block entropy, could also be obtained with a sufficient precision if we had access to an unlimited amount of data. However, data records are necessarily finite and we must resort to other less accurate estimates of the entropy that take into account this finite amount of data. Hereafter, we will use the notation $\hat{a}$ to denote a numerical estimator for an arbitrary random variable $a$. We recall that an unbiased estimator is one for which $\langle \hat a\rangle= a$, while for biased estimators the difference $\langle \hat a\rangle-a$ generally decreases with the sample size $N$.

A first attempt to estimate the entropy of a general sequence $\MS$ is to employ the maximum likelihood estimator (MLE), 
\begin{equation}\label{eq:HMLE}
\hat{H}_{n}^{\MLE} = -\sum_{i=1}^{L^n} \hat{p}^{\MLE}(b_{i}^{(n)})\log\hat{p}^{\MLE}(b_{i}^{(n)}),
\end{equation}
where $\hat{p}^{\MLE}(b_{i}^{(n)}) = \hat{n}_{i}^{(n)}/N_n$ is the relative frequency given by the observed number of occurrences $\hat{n}_{i}^{(n)}$ for the block $b_i^{(n)}$ with respect to the total number $N_n = N-n+1$ of overlapping blocks of size $n$ present in $\MS$.

Despite the fact that $\hat{p}^{\MLE}$ is unbiased, $\hat{H}^{\MLE}_n$ turns out to be biased \cite{Sch_rmann_2004} and this can lead to an extreme underestimation of $H_n$, especially in the undersampled regime $N_n < L^n$ where $\langle \hat{H}^{\MLE}_n\rangle < H_n$. The estimation provided by this method can be considered reliable up to $n_{\text{max}} \sim \log(N)/\log(L)$.

There is not known unbiased estimator for the entropy \cite{paninski} but there is a large number of estimators that manage to improve the results obtained with the MLE \cite{contreras}, allowing one to increase their range of validity as given by $n_{\text{max}}$. In particular, the coverage adjusted estimator proposed by Chao and Shen \cite{chao} provides good results for sequences of independent events.
Here, we will present an estimator that generally improves the coverage adjusted estimator for systems with memory, using a combination of the Horvitz-Thompson adjustment \cite{horvitz} to account for missing elements in the sequence and a correction to the probabilities that takes into account correlations.

\subsection{Horvitz-Thompson correction}

Consider a variable $A(b_i^{(n)})$ that depends on block $b_i^{(n)}$. We are interested in a numerical estimation of the sum of $A$ for all possible blocks, namely $Y_n = \sum_{i=1}^{L^n} A(b_i^{(n)})$. In a finite series $\MS$, not all $L^n$ possible values of $b_i^{(n)}$ will necessarily appear. To account for the missing blocks in the data, Horvitz and Thompson~\cite{horvitz} proposed to estimate $Y_n$ by summing only the contributions of the blocks that do appear in $\MS$ and weighting each term by the probability that the element is included in $\MS$:
\begin{equation}
\hat{Y}_n^{\HT} = \sum_{b_{i}^{(n)}\in \MS}\dfrac{A(b_i^{(n)})}{P(b_i^{(n)}\in \MS)},
\label{eq:HT}
\end{equation}

For a random sequence of length $N$, memory $m=0$ (independent events) and block size $n=1$, the probability of appearance for block $b_i^{(1)}$ can be computed as
\begin{equation}
P(b_i^{(1)} \in \MS) = 1-(1-p(b_{i}^{(1)}))^{N}.
\label{eq:inc_prob}
\end{equation}
In principle, Eq.~\eqref{eq:inc_prob} does not hold in the presence of correlations ($m>0$). It does not hold either for $m=0$ and $n>1$ as the existing overlapping between consecutive blocks already induces correlations in the block series: e.g., in the case $L=2$ and $n=3$ with possible results $z_i=0,1$, the block $(0,0,1)$ can only be followed by the blocks $(0,1,0)$ and $(0,1,1)$. Nevertheless, we have checked in all our numerical simulations that the corrections introduced by these effects are negligible in the limit $N \gg n$ (see \ref{sec:appB}).

Using Eqs.~\eqref{eq:HT} and~\eqref{eq:inc_prob}
in Eq.~\eqref{eq:HBSE}, one arrives at the Horvitz-Thompson estimator for the block entropy
\begin{equation}
\hat{H}_n^{\HT} = -\sum_{b_{i}^{(n)}\in \MS} \dfrac{p(b_{i}^{(n)})\log(p(b_{i}^{(n)}))}{1-(1-p(b_{i}^{(n)}))^{N_n}}.
\label{eq:eHT}
\end{equation}

\subsection{Chao-Shen estimator}

In general, the exact probabilities $p(b_{i}^{(n)})$ that appear in Eq.~\eqref{eq:eHT} are not known. Thus, it is necessary to replace them by their estimators obtained from the sequence $\MS$. The most basic estimator is the maximum-likelihood-estimator $\hat{p}^{\MLE}(b_i^{(n)})$ defined above. However, since not all possible blocks are likely to appear in the finite sequence $\MS$, we note that $\sum_{b_{i}^{(n)}\in \MS} p(b_i^{(n)}) \leq 1$ but, by construction, $\sum_{b_{i}^{(n)}\in \MS} \hat{p}^{\MLE}(b_i^{(n)}) = 1$. Therefore, it is convenient to correct the estimated frequencies before substituting them into Eq.~\eqref{eq:eHT}. Chao and Shen~\cite{chao} proposed to use instead of $\hat{p}^{\MLE}(b_i^{(n)})$ a new estimator $\hat{p}(b_i^{(n)})$ obtained by multiplying $\hat{p}^{\MLE}(b_i^{(n)})$ by the {\slshape sample coverage} $\hat{C}_n$, 
\begin{equation}
\hat{p}(b_i^{(n)}) = \hat{C}_n\hat{p}^{\MLE}(b_i^{(n)}).
\label{eq:corrected_p}
\end{equation}
The new estimator is requested to satisfy the condition $\sum_{b_{i}^{(n)}\in \MS} \hat {p}(b_i^{(n)})=\sum_{b_{i}^{(n)}\in \MS} p(b_i^{(n)})$. From Eq.~\eqref{eq:corrected_p} and using the normalization condition of the MLE estimator, one finds that
\begin{equation}
\hat{C}_n = \sum_{b_{i}^{(n)}\in \MS} p(b_{i}^{(n)})
\label{eq:Cn}
\end{equation}
represents the total probability for the occurrence of the blocks observed in $S$.
$\sum_{b_{i}^{(n)}\in \MS} p(b_{i}^{(n)})<1$ implies that there exist unseen blocks in the data and this is an error source for the estimation of $H_n$.

Again, as the exact values of $p(b_{i}^{(n)})$ are not known, Eq.~\eqref{eq:Cn} can not be used to compute $\hat{C}_n$ for a given sequence $S$. The Good-Turing estimator \cite{good} can be used as an estimate for the sample coverage,
\begin{equation}
\hat{C}^{\GT}_n = 1-\dfrac{N_1^{(n)}}{N_n},
\label{eq:Ccs}
\end{equation}
where $N_1^{(n)}$ is the number of blocks of size $n$ that appear only once in the sequence $\MS$.
Substituting Eq.~\eqref{eq:corrected_p}
into Eq.~\eqref{eq:eHT} and using Eq.~\eqref{eq:Ccs}, one finds the Chao-Shen entropy estimator
\begin{equation}
\hat{H}^{\CS}_n = -\sum_{b_{i}^{(n)}\in \MS} \dfrac{\hat{C}_n^{\GT}\hat{p}^{\MLE}(b_{i}^{(n)})\log(\hat{C}_n^{\GT}\hat{p}^{\MLE}(b_{i}^{(n)}))}{1-(1-\hat{C}_n^{\GT}\hat{p}^{\MLE}(b_{i}^{(n)}))^{N_{n}}},
\label{eq:HCS}
\end{equation}
which has proven to provide very good results for independent sequences. However, we show below that Eq.~\eqref{eq:HCS} does not work so well for correlated data as happens in systems with memory. This is our motivation to present an improved estimator that does take into account correlations.

\subsection{Correlation coverage estimator}

Our proposal follows Eq.~\eqref{eq:corrected_p} but $\hat{C}_n$ is now estimated using a sequential procedure to tackle possible correlations in the sequence. First, we consider the initial part of the sequence, namely $\MS_0=\left(X_1,X_2,\dots,X_{N_1}\right)$, with $N_{1}\equiv n-1+N_n/2$, such that $\MS_0$ contains exactly $N_n/2\equiv N'_n$ blocks of size $n$. We adopt the initial estimator $\hat{C}_n^{(0)}=1$. After observing all the blocks $B^{(n)}_{j=1,\dots,N'_n}$ that appear in $\MS_0$, we take the next observation $B^{(n)}_{+1}$ (we adopt the simplifying notation $B^{(n)}_{+k} \equiv B^{(n)}_{N'_n+k}$, $1\leq k \leq N'_n$), and modify the sample coverage according to
\begin{align}
\begin{split}
\hat{C}^{(1)}_{n} = 
\begin{cases}
\hat{C}^{(0)}_{n} \quad &\text{if }B^{(n)}_{+1} \in \MS_{0}, \\
\hat{C}^{(0)}_{n} - \dfrac{1}{N'_n+1} \quad &\text{if }B^{(n)}_{+1} \not\in \MS_{0}, 
\end{cases}
\end{split}
\end{align}
here the factor $1/(N'_n+1)$ accounts for the probability that the observed block $B^{(n)}_{+1}$ appears for the first time. We repeat the process with the next observation $B^{(n)}_{+2}$ to modify the estimator
\begin{align}
\begin{split}
\hat{C}^{(2)}_{n} = 
\begin{cases}
\hat{C}^{(1)}_{n} \quad &\text{if }B^{(n)}_{+2} \in \MS_{+1}, \\
\hat{C}^{(1)}_{n} - \dfrac{1}{N'_n+2} \quad &\text{if }B^{(n)}_{+2} \not\in \MS_{+1}, 
\end{cases}
\end{split}
\end{align}
where $\MS_{+i} = \MS_{0} \cup (X_{N_1+1}, \ldots,X_{N_1+i})$.
At variance with the coverage given by Eq.~\eqref{eq:Ccs}, which is obtained from a single observation from the whole series, this procedure has the advantage of iteratively updating $\hat{C}_n^{(k)}$, $k=0,1,2,\dots,N'_n$ based on the previously observed data. We continue with this procedure until we arrive at the final value of the 
{\slshape correlation coverage} (CC) estimator
\begin{equation}
\hat{C}_n^{\CC} \equiv \hat{C}^{\left(N'_n\right)}_{n} = 1 - \sum_{j=1}^{N'_n} \dfrac{1}{{N'_n+j}}I \left(B^{(n)}_{+j}\not\in \MS_{+(j-1)} \right),
\end{equation}
where the indicator function $I(Z)$ yields $1$ if the event $Z$ is true and $0$ otherwise.

Finally, we substitute the corrected probabilities into Eq.~\eqref{eq:eHT} to obtain the correlation coverage-adjusted entropy estimator
\begin{equation}\label{eq:Hnew}
\hat{H}^{\CC}_n = -\sum_{b_{i}^{(n)}\in \MS} \dfrac{\hat{C}_n^{\CC}\hat{p}^{\MLE}(b_{i}^{(n)})\log(\hat{C}_n^{\CC}\hat{p}^{\MLE}(b_{i}^{(n)}))}{1-(1-\hat{C}_n^{\CC}\hat{p}^{\MLE}(b_{i}^{(n)}))^{N_{n}}}.
\end{equation}

\begin{figure}[t]
\includegraphics[width=\columnwidth]{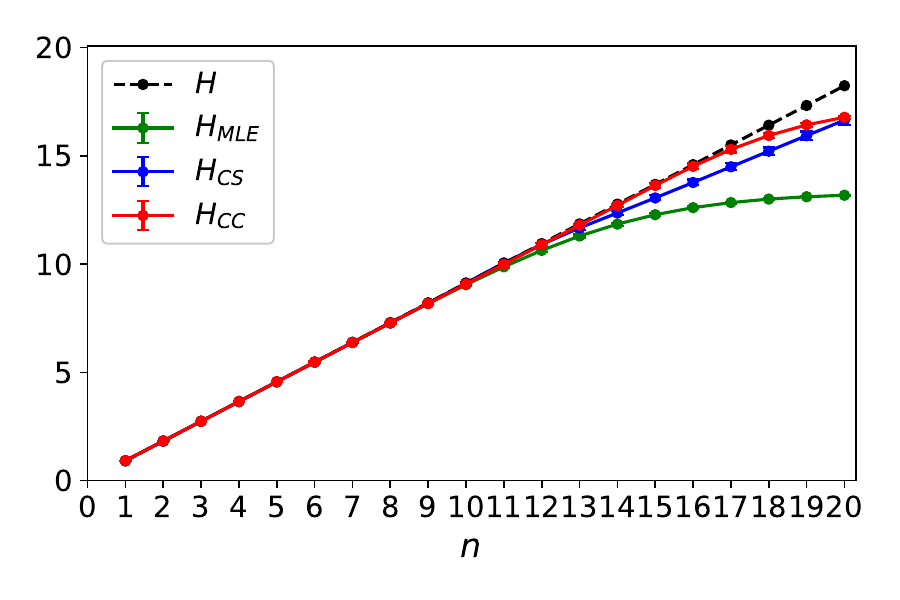}
\caption{Exact Shannon entropy per block of size $n$ (dotted line) for a particular case of a Markovian binary system with fixed transition probabilities $p(0|0)=0.7$ and $p(1|1) = 0.6$. A sequence of $N=10^4$ realizations is numerically generated from which we calculate three different entropy estimators discussed in the main text. The best performance is shown for the estimator that takes into account correlations (red dots).}
\label{fig:Hvsn}
\end{figure}

As an example, in Fig.~\ref{fig:Hvsn} we show the exact entropy $H_n$ (dashed black line and dots) as a function of the block size for a Markovian ($m=1$) binary process that takes the values $z=0,1$, with transition probabilities $P_0\equiv p(0|0)= P(X_s=0|X_{s-1}=0)=0.7$ and $P_1\equiv p(1|1) = P(X_s=1|X_{s-1}=1)= 0.6$. Note that, as expected, $H_n$ is linear from $n\ge 1$. To compare with, we also depict the results obtained with the maximum-likelihood estimators given by Eq.~\eqref{eq:HMLE} (green line and dots), the Chao-Shen estimator of Eq.~\eqref{eq:HCS} (blue line and dots) and our coverage-adjusted estimator proposed in Eq.~\eqref{eq:Hnew} (red line and dots), all calculated from a sequence of $N=10^4$ elements generated numerically. Notably, the proposed estimator performs better than both the MLE and Chao-Shen entropies and perfectly agrees with the exact result for a wide range of block sizes up to $n\lesssim 17$ while the MLE and the Chao-Shen estimators deviate earlier from the exact $H_n$. In fact, the MLE estimator provides a good approximation up to a value of $n\lesssim 12$ which is close to the expected limit $n\sim\log(N)/\log(L)\sim 13$, while the validity of the Chao-Shen estimator extends up to $n\lesssim 13$ and our estimator even further.
In \ref{sec:appC} we present a similar comparison for sequences with memory $2$ and $3$.

Departures from the exact value will show up only in the extremely undersampled regime ($N_n \ll L^n$). In this case, every observation will constitute a different element. As a consequence, the sum in Eq.~\eqref{eq:Hnew} will include all blocks in the second part and $\hat{C}^{\CC}_n =1-\sum_{j=1}^{N'_n}\frac{1}{N'_n+j}\simeq 1-\ln(2)\simeq 0.307$, for large $N$. The proximity of the computed estimator to this limiting value indicates a largely undersampled sequence and determines the validity of the estimators to the coverage $C_n$.

\begin{figure}[t]
\begin{center}
{\includegraphics[width=.65\columnwidth]{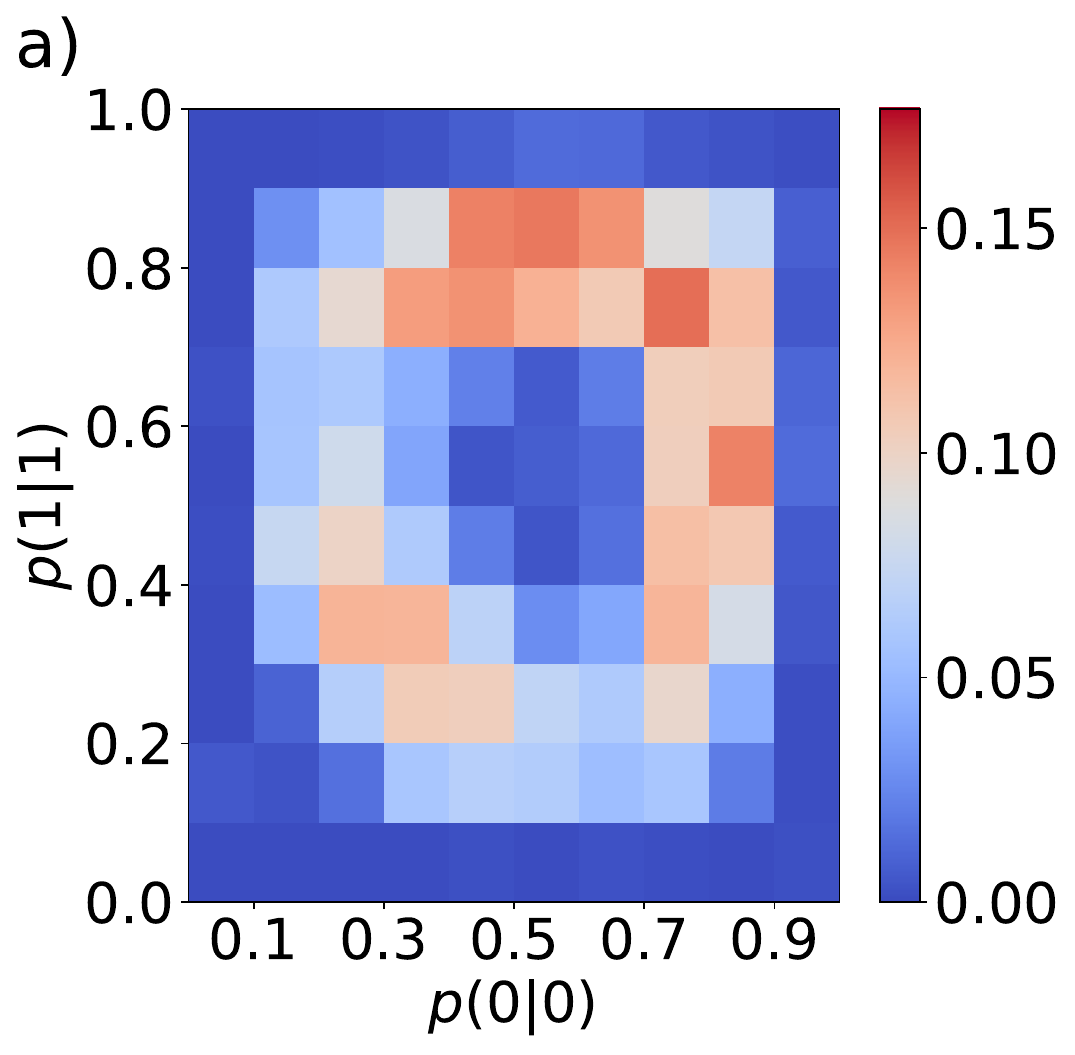}}\quad
{\includegraphics[width=.65\columnwidth]{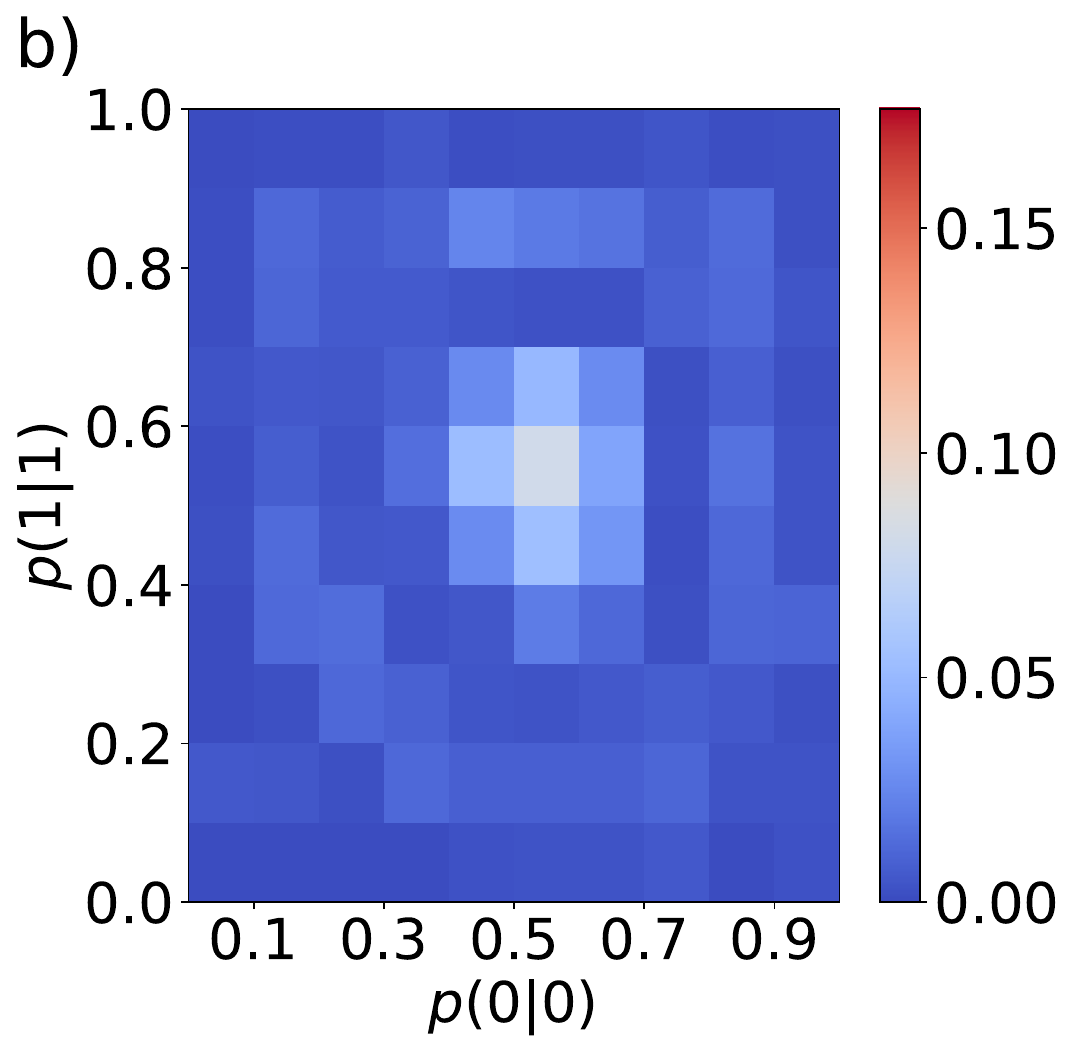}}
\end{center}
\caption{Comparison between the Chao-Shen entropy, panel (a), and our proposed estimator for correlated systems, panel (b). Calculations are done for a binary system with memory $m = 1$ and varying transition probabilities $p(0|0)\equiv P_0$ and $p(1|1)\equiv P_1$. Colors represent the departure of each estimator (computed from $10^4$ realizations) from the exact entropy, as measured by the mean squared error $\epsilon(P_0,P_1)\equiv\frac{1}{n_{\text{max}}}\sum_{n=1}^{n_{\text{max}}} (H_n-\hat{H}_n)^2$ with $n_\text{max}=17$. Adding all values for $\epsilon(P_0, P_1)$ using the grid size $\Delta P = 0.1$, we find that the overall error reaches a value of $4.65$ in (a) but only of $0.90$ in (b).}
\label{fig:colormaps}
\end{figure}

For a more detailed comparison, we show in panel (a) of Fig.~\ref{fig:colormaps} the performance of the Chao-Shen estimator given by Eq.~\eqref{eq:HCS}, and in panel (b) that of the our correlation estimator from Eq.~\eqref{eq:Hnew}, both calculated for a process with $m=1$ and $L=2$. We show the results as a function of all possible combinations of transition probabilities $P_0$ and $P_1$, the remaining probabilities being derived from the normalization conditions $p(1|0)=1-P_0$ and $p(0|1)=1-P_1$. We measure the goodness of each estimator for particular values of $P_0$ and $P_1$ with the mean squared error $\epsilon(P_0,P_1)\equiv\frac{1}{n_{\text{max}}}\sum_{n=1}^{n_{\text{max}}} (H_n-\hat{H}_n)^2$. To improve the statistics of the error, for each set of transition probabilities $(P_0,P_1)$ we generate $M=20$ series, each with $N=10^4$ elements, and average the mean squared error over the $M$ values. We use in all cases $n_{\text{max}}=17$ and plot the resulting values in a color code. As shown in the figure, quite generally, Eq.~\eqref{eq:Hnew} performs better than Eq.~\eqref{eq:HCS}, with the possible exception of the cases near $P_0,P_1\approx 1/2$ (the independent case with equal outcome probability), in which the latter is slightly better. The overall performance is obtained by adding all values of $\epsilon(P_0,P_1)$ using a grid size $\Delta P=0.1$ in Fig.~\ref{fig:colormaps}, resulting in an aggregated mean squared error five times larger for the Chao-Shen estimator as compared with the correlation coverage-adjusted estimator.

\section{Determination of the memory of a sequence}
\label{sec:results}

Having demonstrated the usefulness of the estimator given
by Eq.~\eqref{eq:Hnew} for Markovian systems,
we now return to the method of Sec.~\ref{sec:method} for the determination
of the memory in discrete sequences.

Suppose we are given a finite time series $\cal{S}$ that describes a phenomenon for which we would like to determine its memory $m$. As explained above, we will use the criterion given by Eq.~\eqref{eq:m} after computing $\Delta_\mu$ following Eqs.~(\ref{eq:Hmu},\ref{eq:Deltamu}) and using a suitable estimator $\hat{H}_n$ for the block entropy $H_n$. 
For a given entropy estimator with a known $n_{\text{max}}$ and fixed $N$ and $L$, a meaningful linear fit for the calculation of $\Delta_{\mu}$ as given by Eq.~\eqref{eq:Deltamu} requires that $\mu \leq n_{\text{max}}-2$
(at least three points are required for a meaningful fit to a straight line).
Since the chosen entropy estimator works for block sizes up to $n_{\text{max}}$, our method can provide in principle an accurate result if the system under study has memory $m \leq n_{\text{max}}-2$ but would fail otherwise.

In Figs.~\ref{fig:Hvsn} and \ref{fig:m=2,3} we have shown evidence that $n_{\text{max}}^{\text{\tiny{CC}}} > n_{\text{max}}^{\text{\tiny{MLE}}}$. Hence, there exist three regimes for the estimation of the memory $m$: i) if $m \leq n_{\text{max}}^{\text{\tiny{MLE}}}-2$, the fit to a straight line works with both MLE and CC entropy estimators; ii) if $n_{\text{max}}^{\text{\tiny{MLE}}}-2 < m \leq n_{\text{max}}^{\text{\tiny{CC}}}-2$, the MLE estimator is insufficient; iii) if $m > n_{\text{max}}^{\text{\tiny{CC}}}-2$ the memory is exceedingly large and both estimators are inadequate.

It is worth mentioning that, in general, the value of $n_{\text{max}}$ for a certain estimator is not the same for every sequence since $n_{\text{max}}$ may depend on the transition probabilities. To be conservative, in this section we use $n_{\text{max}}^{\text{\tiny{CC}}} = n_{\text{max}}^{\text{\tiny{MLE}}}$ even though our estimator certainly provides good results for larger block sizes.

Due to the limitations of all estimators we expect that $\Delta_\mu$ is determined within an error that must be taken into account. To do so, we consider $M$ series of data ${\cal S}^{(i)},\, i=1,\dots,M$, all with the same length, obtained either by repeating the experiment $M$ times or by splitting the original series in $M$ disconnected sequences. For each sequence we calculate the corresponding value of $\Delta_\mu^{(i)}$. Then, we calculate the mean $\bar \Delta_\mu$ and standard deviation $\sigma_\mu$ for the obtained values $\Delta_\mu^{(i)}$. The condition given by Eq.~\eqref{eq:m} is transformed into the criterion that the mean value $\bar \Delta_\mu$ is consistent with the value $0$ within the standard deviation,
\begin{equation}\label{eq:mmu}
m=\text{min}(\mu: \bar{\Delta}_{\mu}-\sigma_\mu\le 0).
\end{equation}

\subsection{Numerical simulation}
\label{sec:results-numerical}

\begin{figure}[t]
\centering
{\includegraphics[width=.65\columnwidth]{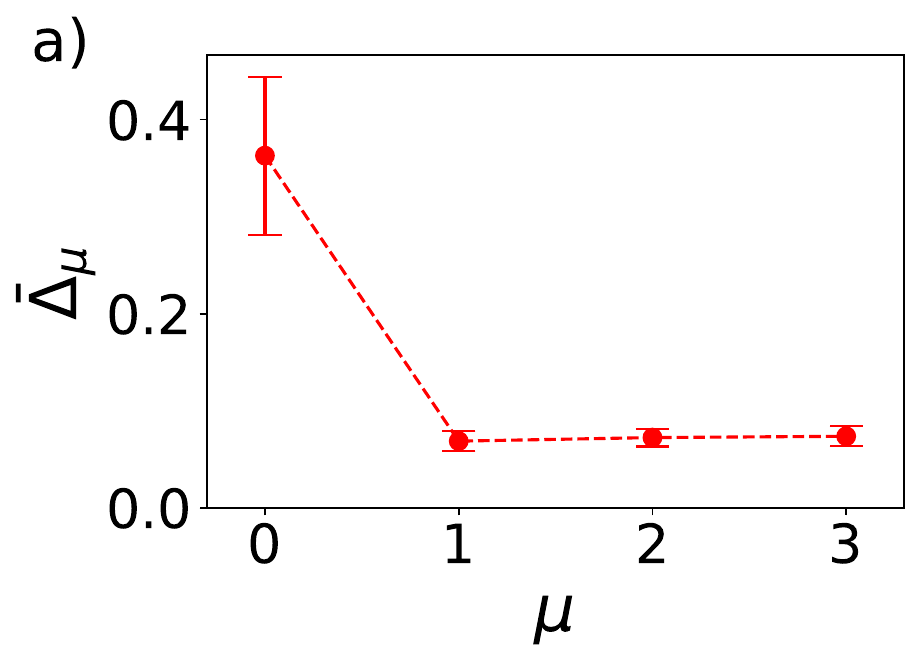}}\quad
{\includegraphics[width=.65\columnwidth]{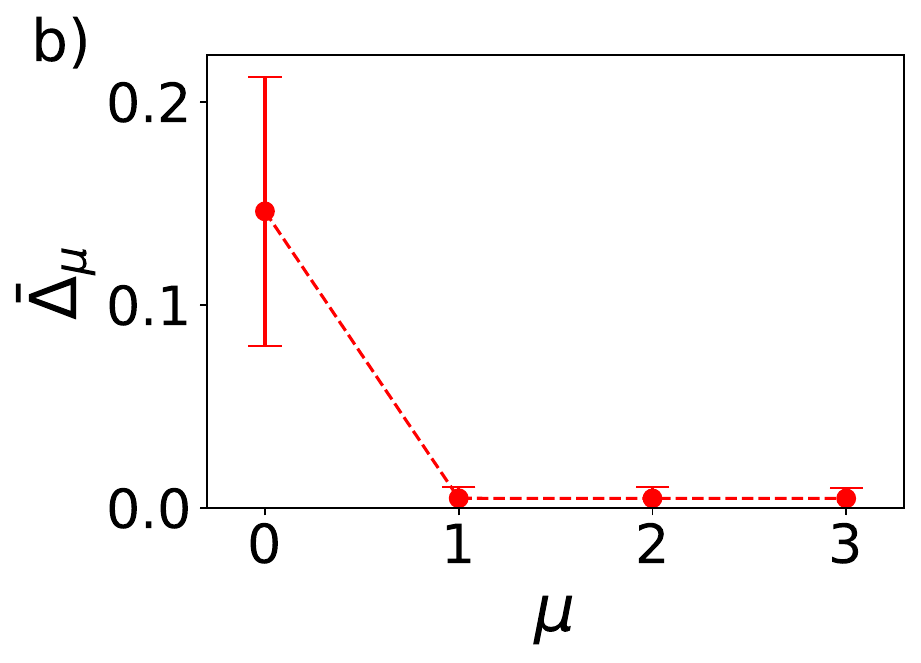}}
\caption{Mean squared error $\bar\Delta_\mu$ as a function of the trial entropy with $n_\text{max}=10$ after averaging over $M=20$ realizations of a Markov chain with $N=1000$ data points, memory $m=1$ and transition probabilities $p(0|0)=0.7$ and $p(1|1) = 0.6$. In panel a) we use the MLE estimator while panel b) is generated with the correlation coverage-adjusted estimator. Error bars are given by the standard deviation. Importantly, our proposed estimator clearly yields the memory value $m=1$ applying Eq.~\eqref{eq:mmu}.}
\label{fig:method2}
\end{figure}

We first present results arising from controlled numerical simulations. We generate $M=20$ series of length $N=1000$ with memory $m=1$ and possible values $z=0,1$ ($L=2$). We show only results corresponding to $p(0|0)=0.7$ and $p(1|1) = 0.6$ but similar conclusions are obtained quite generally for different values of the transition probabilities. 
In Fig.~\ref{fig:method2} we plot the values of $\bar\Delta_\mu$ obtained using the method explained above. In Fig.~\ref{fig:method2}(a) we employ the MLE estimator for the block entropy, while Fig.~\ref{fig:method2}(b) uses the correlation coverage-adjusted estimator. Remarkably, the MLE entropy is not able to yield any useful result, and it is thus not possible to determine the sequence memory since Eq.~\eqref{eq:mmu} is never satisfied. In stark contrast, the calculation of $\bar\Delta_\mu$ and $\sigma_\mu$ using the entropy estimator given by Eq.~\eqref{eq:Hnew} clearly shows that the criterion Eq.~\eqref{eq:mmu} yields the correct result $m=1$.

\begin{figure}[t]
\centering
{\includegraphics[width=.65\columnwidth]{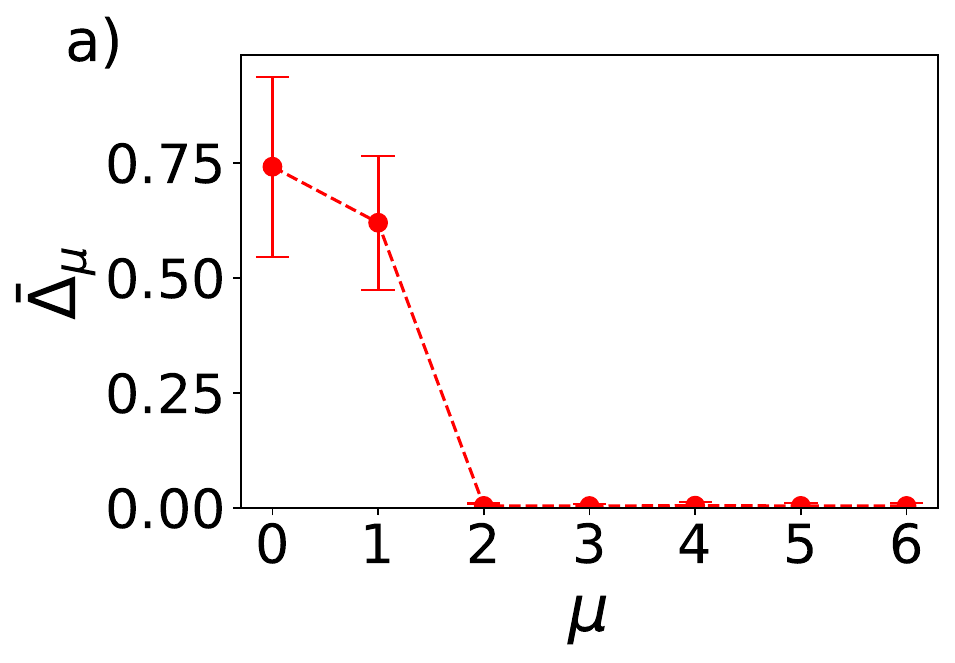}}\quad
{\includegraphics[width=.65\columnwidth]{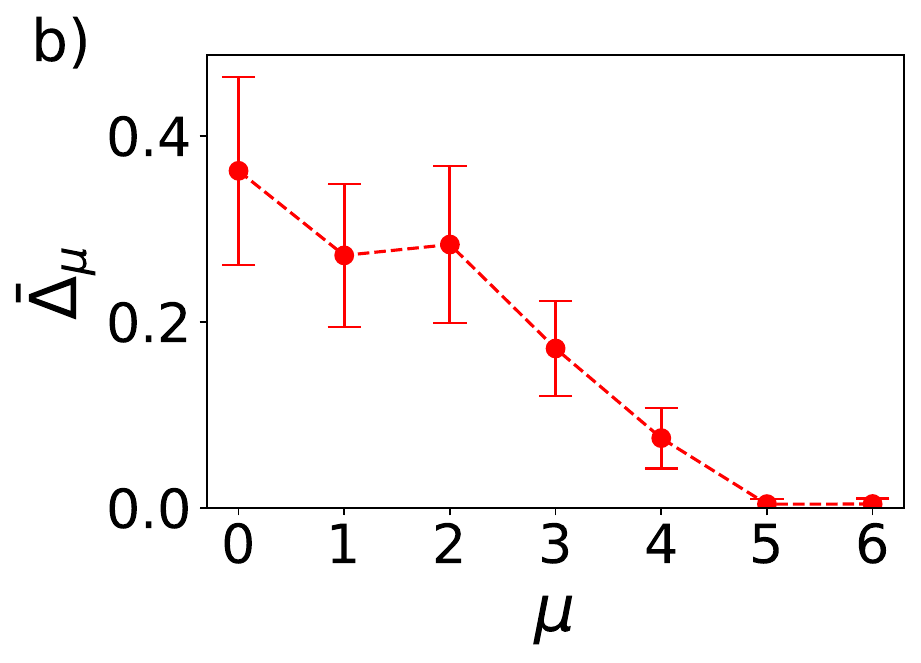}}
\caption{Mean squared error $\bar\Delta_\mu$ as a function of the trial entropy with $n_\text{max}=10$ after averaging over $M=20$ realizations of a Markov chain with $N=1000$ data points, transition probabilities chosen randomly from an uniform distribution in the interval $(0,1)$ and memory $m=2$ panel a), and $m=5$, panel b), using the method explained in the main text with correlation coverage-adjusted estimator. Error bars are given by the standard deviation. Importantly, in both cases, our proposed entropy estimator clearly yields the right memory value applying Eq.~\eqref{eq:mmu}.}
\label{fig:method3}
\end{figure}

In Fig.~\ref{fig:method3}(a) and (b) we present results arising from similar numerical simulations as before but now generating sequences with memory $m=2$ and $m=5$, respectively (same values of $M$, $N$ and $L$ as before) and transition probabilities chosen randomly from an uniform distribution in the interval $(0,1)$. We plot the values of $\bar\Delta_\mu$ obtained using the correlation coverage-adjusted estimator for the entropy. Again, the usage of our proposed entropy estimator allows us to accurately determine the memory of the system.
Similar conclusions are generally obtained for a different set of transition probabilities.

These numerical experiments illustrate the importance of having a reliable entropy estimator to successfully apply the method of memory determination. 

\subsection{Daily precipitations}
\label{sec:results-daily}
We have thus far tested our proposed method only with numerically generated sequences of known memories.
We now test the method with real data. It has been widely accepted that the occurrence (or not) of precipitation can be modeled as a system of memory $1$, but it is also well known that this assumption has several shortcomings \cite{doi:10.3137/ao.430102}. There have been some attempts to improve this model by studying sequences of daily data worldwide. It has been found~\cite{WilsonKemsley2021} that the model memory depends on the geographical location. Here, we select a few locations and compare the results with those obtained using the BIC method~\cite{bic}.

We collect data from the Global Historical Climatology Network Daily~\cite{men12}. 
For each location, we record the available observation for daily precipitation, setting the threshold at $0.1$~mm to specify if a day is rainy or not. This way we produce a time series with two states ($L=2)$ and length $N_s \lesssim 25000$ (the exact value of $N_s$ varies for each location). 
Then, this series is divided in $M=5$ sequences of equal lengths $N = N_s/5$ and for each of this sequence we estimate the block entropies $\hat{H}_{n}^{\CC}$ for $n = 1,\ldots,12$, from which we obtain the mean values $\bar{\Delta}_{\mu}$ and their standard deviations $\sigma_{\mu}$. It should be noted that the use of the correlation coverage estimator for the entropy allows us to obtain reliable results up to $n_\text{max} = 12$. If we had used the MLE estimator, this limit value would have been $n_\text{max}\sim 11$.

\begin{figure}
\subfloat{\includegraphics[width=0.48\columnwidth]{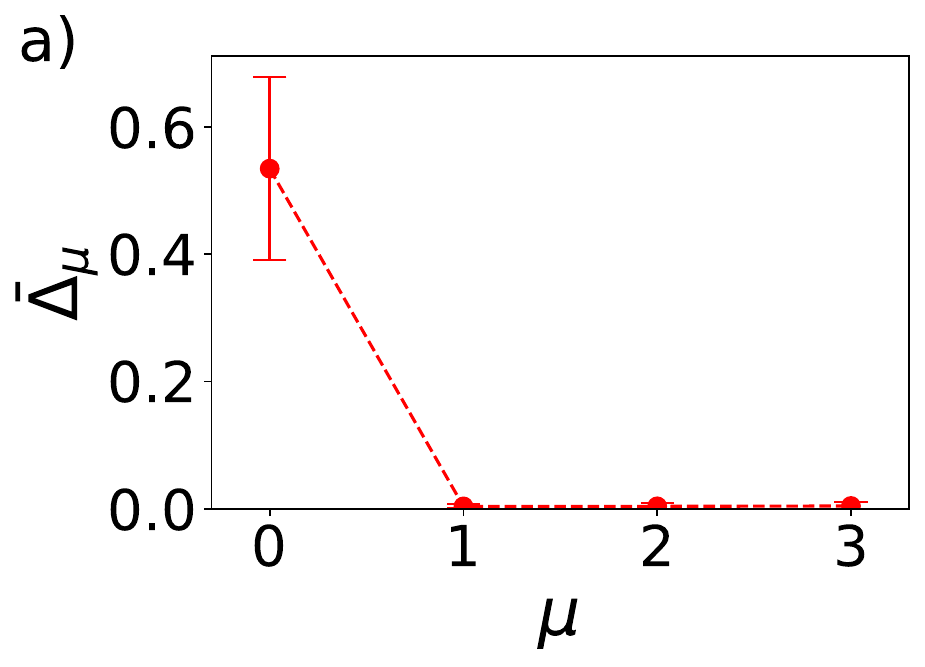}}
 \label{subfig:rome}\hfill
\subfloat{\includegraphics[width=0.48\columnwidth]{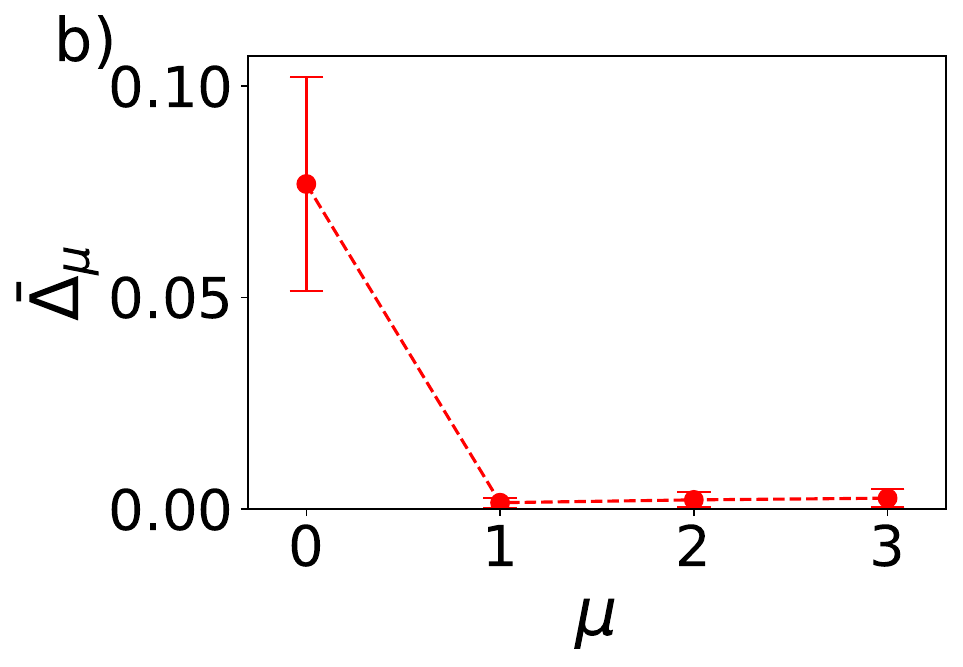}}
 \label{subfig:dallas}
\newline
 \subfloat{\includegraphics[width=0.48\columnwidth]{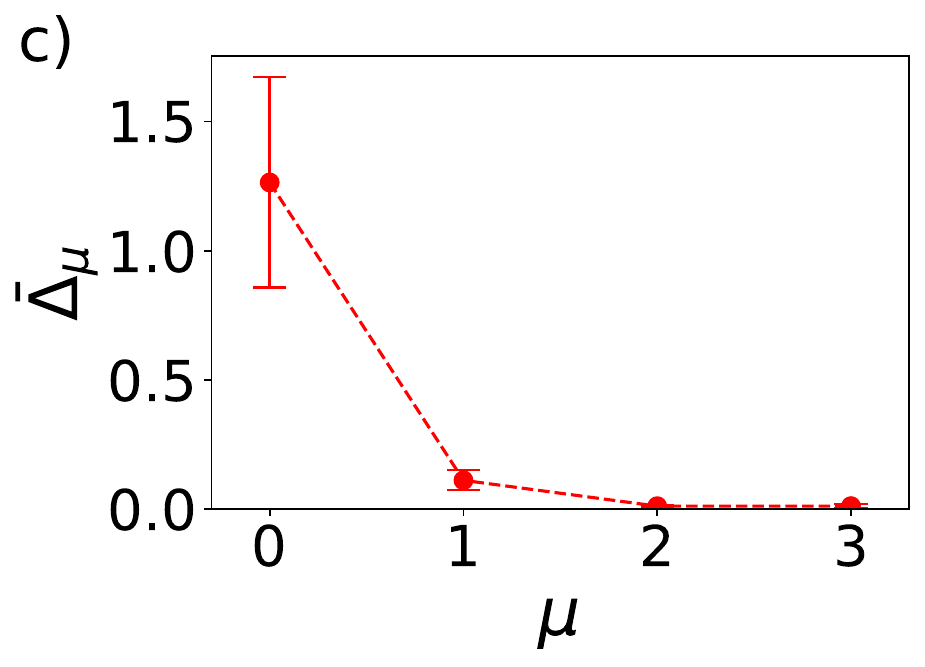}}
 \label{subfig:thailand}
\hfill
 \subfloat{\includegraphics[width=0.48\columnwidth]{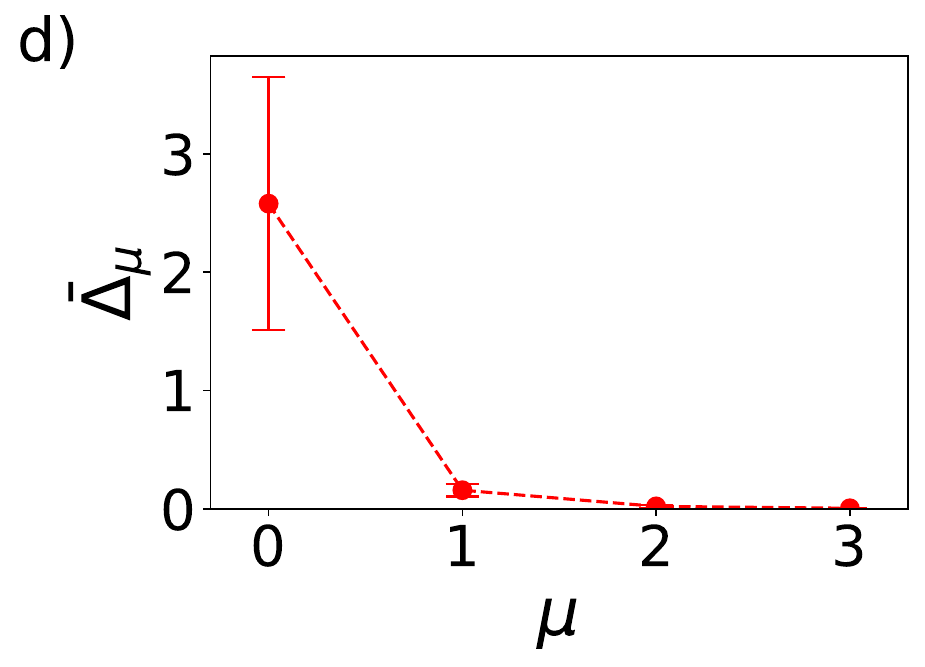}}
 \label{subfig:myanmar}
 \caption{Results of the method explained in Section \ref{sec:method} applied to sequences of data of daily precipitations for four locations: (a) Rome-Italy, (b) Dallas-USA, (c) Bangkok-Thailand and (d) Than Lwin-Myanmar.} 
 \label{fig:precipitations}
\end{figure}

In Fig.~\ref{fig:precipitations} we show the results obtained for four locations: Rome, Dallas, Bangkok and Than Lwin. For the first two locations, our method predicts $m=1$ whereas for the second two places the procedure suggests that both series are better described with $m=2$. These results are in excellent agreement with the BIC method (see Fig.~5 in Ref.~\cite{WilsonKemsley2021}) and consequently validate our technique
for the memory determination in real data sequences
modeled with Markov chains of arbitrary order.

\vspace{10pt}
\section{Conclusions}
\label{sec:conclusions}
We have developed a novel method to determine the memory of a discrete sequence. Importantly, the method is valid for both Markovian and non-Markovian systems. Since the technique relies on the calculation of the Shannon entropy as a function of the block size, it is crucial to additionally propose an entropy estimator that gives good results for correlated systems. To this end, we have introduced a correction to the estimated probabilities to amend the source of error stemming from unseen elements in small samples.
Our estimator is shown to significantly increase the accuracy of the entropy for systems with memory. Both numerically generated sequences and real data series have been used as benchmarks. These successful results will certainly encourage further applications of the proposals discussed in this work.
It is left as a future project to present a more detailed comparison of our entropy estimator with different estimators that have shown promising results for independent sequences~\cite{nsb, archer, shr, grassberger}.

\section*{Acknowledgements}
\label{sec:Acknowledgements}
Partial financial support has been received from MCIN/AEI/10.13039/501100011033 and the Fondo Europeo de Desarrollo Regional (FEDER, UE) through project PACSS (RTI2018-093732-B-C21), APASOS (PID2021-122256NB-C21) and the Mar{\'\i}a de Maeztu Program for units of Excellence in R\&D, grant MDM-2017-0711.

\bibliography{references}

\providecommand{\noopsort}[1]{}\providecommand{\singleletter}[1]{#1}%
\begin{thebibliography}{39}%
\makeatletter
\providecommand \@ifxundefined [1]{%
 \@ifx{#1\undefined}
}%
\providecommand \@ifnum [1]{%
 \ifnum #1\expandafter \@firstoftwo
 \else \expandafter \@secondoftwo
 \fi
}%
\providecommand \@ifx [1]{%
 \ifx #1\expandafter \@firstoftwo
 \else \expandafter \@secondoftwo
 \fi
}%
\providecommand \natexlab [1]{#1}%
\providecommand \enquote  [1]{``#1''}%
\providecommand \bibnamefont  [1]{#1}%
\providecommand \bibfnamefont [1]{#1}%
\providecommand \citenamefont [1]{#1}%
\providecommand \href@noop [0]{\@secondoftwo}%
\providecommand \href [0]{\begingroup \@sanitize@url \@href}%
\providecommand \@href[1]{\@@startlink{#1}\@@href}%
\providecommand \@@href[1]{\endgroup#1\@@endlink}%
\providecommand \@sanitize@url [0]{\catcode `\\12\catcode `\$12\catcode
  `\&12\catcode `\#12\catcode `\^12\catcode `\_12\catcode `\%12\relax}%
\providecommand \@@startlink[1]{}%
\providecommand \@@endlink[0]{}%
\providecommand \url  [0]{\begingroup\@sanitize@url \@url }%
\providecommand \@url [1]{\endgroup\@href {#1}{\urlprefix }}%
\providecommand \urlprefix  [0]{URL }%
\providecommand \Eprint [0]{\href }%
\providecommand \doibase [0]{http://dx.doi.org/}%
\providecommand \selectlanguage [0]{\@gobble}%
\providecommand \bibinfo  [0]{\@secondoftwo}%
\providecommand \bibfield  [0]{\@secondoftwo}%
\providecommand \translation [1]{[#1]}%
\providecommand \BibitemOpen [0]{}%
\providecommand \bibitemStop [0]{}%
\providecommand \bibitemNoStop [0]{.\EOS\space}%
\providecommand \EOS [0]{\spacefactor3000\relax}%
\providecommand \BibitemShut  [1]{\csname bibitem#1\endcsname}%
\let\auto@bib@innerbib\@empty
\bibitem [{\citenamefont {Cox}\ and\ \citenamefont {Miller}(2017)}]{cox17}%
  \BibitemOpen
  \bibfield  {author} {\bibinfo {author} {\bibfnamefont {David~Roxbee}\
  \bibnamefont {Cox}}\ and\ \bibinfo {author} {\bibfnamefont {Hilton~David}\
  \bibnamefont {Miller}},\ }\href@noop {} {\emph {\bibinfo {title} {The theory
  of stochastic processes}}}\ (\bibinfo  {publisher} {Routledge},\ \bibinfo
  {year} {2017})\BibitemShut {NoStop}%
\bibitem [{\citenamefont {Markov}(2006)}]{markov}%
  \BibitemOpen
  \bibfield  {author} {\bibinfo {author} {\bibfnamefont {A.~A.}\ \bibnamefont
  {Markov}},\ }\bibfield  {title} {\enquote {\bibinfo {title} {An example of
  statistical investigation of the text {Eugene Onegin} concerning the
  connection of samples in chains},}\ }\href {\doibase
  10.1017/S0269889706001074} {\bibfield  {journal} {\bibinfo  {journal}
  {Science in Context}\ }\textbf {\bibinfo {volume} {19}},\ \bibinfo {pages}
  {591--600} (\bibinfo {year} {2006})}\BibitemShut {NoStop}%
\bibitem [{\citenamefont {Gardiner}(1965)}]{gar04}%
  \BibitemOpen
  \bibfield  {author} {\bibinfo {author} {\bibfnamefont {C.~W.}\ \bibnamefont
  {Gardiner}},\ }\href@noop {} {\emph {\bibinfo {title} {Handbook of stochastic
  methods for Physics, Chemistry and the Natural Sciences}}}\ (\bibinfo
  {publisher} {Springer},\ \bibinfo {address} {Berlin},\ \bibinfo {year}
  {1965})\BibitemShut {NoStop}%
\bibitem [{\citenamefont {Churchill}(1989)}]{chu89}%
  \BibitemOpen
  \bibfield  {author} {\bibinfo {author} {\bibfnamefont {Gary~A.}\ \bibnamefont
  {Churchill}},\ }\bibfield  {title} {\enquote {\bibinfo {title} {Stochastic
  models for heterogeneous {DNA} sequences},}\ }\href {\doibase
  10.1016/S0092-8240(89)80049-7} {\bibfield  {journal} {\bibinfo  {journal}
  {Bulletin of Mathematical Biology}\ }\textbf {\bibinfo {volume} {51}},\
  \bibinfo {pages} {79--94} (\bibinfo {year} {1989})}\BibitemShut {NoStop}%
\bibitem [{\citenamefont {Wilks}\ and\ \citenamefont {Wilby}(1999)}]{wil99}%
  \BibitemOpen
  \bibfield  {author} {\bibinfo {author} {\bibfnamefont {D.~S.}\ \bibnamefont
  {Wilks}}\ and\ \bibinfo {author} {\bibfnamefont {R.~L.}\ \bibnamefont
  {Wilby}},\ }\bibfield  {title} {\enquote {\bibinfo {title} {The weather
  generation game: a review of stochastic weather models},}\ }\href {\doibase
  10.1177/030913339902300302} {\bibfield  {journal} {\bibinfo  {journal}
  {Progress in Physical Geography: Earth and Environment}\ }\textbf {\bibinfo
  {volume} {23}},\ \bibinfo {pages} {329--357} (\bibinfo {year}
  {1999})}\BibitemShut {NoStop}%
\bibitem [{\citenamefont {Page}\ \emph {et~al.}(1999)\citenamefont {Page},
  \citenamefont {Brin}, \citenamefont {Motwami},\ and\ \citenamefont
  {Winograd}}]{pag06}%
  \BibitemOpen
  \bibfield  {author} {\bibinfo {author} {\bibfnamefont {Lawrence}\
  \bibnamefont {Page}}, \bibinfo {author} {\bibfnamefont {Sergey}\ \bibnamefont
  {Brin}}, \bibinfo {author} {\bibfnamefont {Rajeev}\ \bibnamefont {Motwami}},
  \ and\ \bibinfo {author} {\bibfnamefont {Terry}\ \bibnamefont {Winograd}},\
  }\href@noop {} {\emph {\bibinfo {title} {The PageRank citation ranking:
  Bringing order to the Web}}},\ \bibinfo {type} {Technical Report}\ \bibinfo
  {number} {1999-66}\ (\bibinfo  {institution} {Stanford InfoLab},\ \bibinfo
  {year} {1999})\BibitemShut {NoStop}%
\bibitem [{\citenamefont {H{\"a}nggi}\ and\ \citenamefont
  {Thomas}(1982)}]{han82}%
  \BibitemOpen
  \bibfield  {author} {\bibinfo {author} {\bibfnamefont {Peter}\ \bibnamefont
  {H{\"a}nggi}}\ and\ \bibinfo {author} {\bibfnamefont {Harry}\ \bibnamefont
  {Thomas}},\ }\bibfield  {title} {\enquote {\bibinfo {title} {Stochastic
  processes: Time evolution, symmetries and linear response},}\ }\href
  {\doibase 10.1016/0370-1573(82)90045-X} {\bibfield  {journal} {\bibinfo
  {journal} {Physics Reports}\ }\textbf {\bibinfo {volume} {88}},\ \bibinfo
  {pages} {207--319} (\bibinfo {year} {1982})}\BibitemShut {NoStop}%
\bibitem [{\citenamefont {Mayzelis}\ \emph {et~al.}(2007)\citenamefont
  {Mayzelis}, \citenamefont {Apostolov}, \citenamefont {Melnyk}, \citenamefont
  {Usatenko},\ and\ \citenamefont {Yampol'skii}}]{may07}%
  \BibitemOpen
  \bibfield  {author} {\bibinfo {author} {\bibfnamefont {ZA}~\bibnamefont
  {Mayzelis}}, \bibinfo {author} {\bibfnamefont {SS}~\bibnamefont {Apostolov}},
  \bibinfo {author} {\bibfnamefont {SS}~\bibnamefont {Melnyk}}, \bibinfo
  {author} {\bibfnamefont {OV}~\bibnamefont {Usatenko}}, \ and\ \bibinfo
  {author} {\bibfnamefont {VA}~\bibnamefont {Yampol'skii}},\ }\bibfield
  {title} {\enquote {\bibinfo {title} {{Additive N-step Markov chains as
  prototype model of symbolic stochastic dynamical systems with long-range
  correlations}},}\ }\href {\doibase 10.1016/j.chaos.2007.01.054} {\bibfield
  {journal} {\bibinfo  {journal} {Chaos, Solitons \& Fractals}\ }\textbf
  {\bibinfo {volume} {34}},\ \bibinfo {pages} {112--128} (\bibinfo {year}
  {2007})}\BibitemShut {NoStop}%
\bibitem [{\citenamefont {Raftery}(1985)}]{raf85}%
  \BibitemOpen
  \bibfield  {author} {\bibinfo {author} {\bibfnamefont {Adrian~E.}\
  \bibnamefont {Raftery}},\ }\bibfield  {title} {\enquote {\bibinfo {title} {A
  model for high-order {Markov} chains},}\ }\href {\doibase
  10.1111/j.2517-6161.1985.tb01383.x} {\bibfield  {journal} {\bibinfo
  {journal} {Journal of the Royal Statistical Society: Series B
  (Methodological)}\ }\textbf {\bibinfo {volume} {47}},\ \bibinfo {pages}
  {528--539} (\bibinfo {year} {1985})}\BibitemShut {NoStop}%
\bibitem [{\citenamefont {Yulmetyev}\ \emph {et~al.}(2006)\citenamefont
  {Yulmetyev}, \citenamefont {Demin}, \citenamefont {Panischev}, \citenamefont
  {H{\"a}nggi}, \citenamefont {Timashev},\ and\ \citenamefont
  {Vstovsky}}]{yul06}%
  \BibitemOpen
  \bibfield  {author} {\bibinfo {author} {\bibfnamefont {Renat~M}\ \bibnamefont
  {Yulmetyev}}, \bibinfo {author} {\bibfnamefont {Sergey~A}\ \bibnamefont
  {Demin}}, \bibinfo {author} {\bibfnamefont {O~Yu}\ \bibnamefont {Panischev}},
  \bibinfo {author} {\bibfnamefont {Peter}\ \bibnamefont {H{\"a}nggi}},
  \bibinfo {author} {\bibfnamefont {Serge~F}\ \bibnamefont {Timashev}}, \ and\
  \bibinfo {author} {\bibfnamefont {Grigoriy~V}\ \bibnamefont {Vstovsky}},\
  }\bibfield  {title} {\enquote {\bibinfo {title} {{Regular and stochastic
  behavior of Parkinsonian pathological tremor signals}},}\ }\href {\doibase
  10.1016/j.physa.2006.01.077} {\bibfield  {journal} {\bibinfo  {journal}
  {Physica A: Statistical Mechanics and its Applications}\ }\textbf {\bibinfo
  {volume} {369}},\ \bibinfo {pages} {655--678} (\bibinfo {year}
  {2006})}\BibitemShut {NoStop}%
\bibitem [{\citenamefont {Ho}\ and\ \citenamefont {Cao}(2012)}]{ho12}%
  \BibitemOpen
  \bibfield  {author} {\bibinfo {author} {\bibfnamefont {Dung~T}\ \bibnamefont
  {Ho}}\ and\ \bibinfo {author} {\bibfnamefont {Tru~H}\ \bibnamefont {Cao}},\
  }\bibfield  {title} {\enquote {\bibinfo {title} {{A high-order hidden Markov
  model for emotion detection from textual data}},}\ }in\ \href@noop {} {\emph
  {\bibinfo {booktitle} {Pacific rim knowledge acquisition workshop}}}\
  (\bibinfo {organization} {Springer},\ \bibinfo {year} {2012})\ pp.\ \bibinfo
  {pages} {94--105}\BibitemShut {NoStop}%
\bibitem [{\citenamefont {Seifert}\ \emph {et~al.}(2012)\citenamefont
  {Seifert}, \citenamefont {Gohr}, \citenamefont {Strickert},\ and\
  \citenamefont {Grosse}}]{sei12}%
  \BibitemOpen
  \bibfield  {author} {\bibinfo {author} {\bibfnamefont {Michael}\ \bibnamefont
  {Seifert}}, \bibinfo {author} {\bibfnamefont {Andr{\'e}}\ \bibnamefont
  {Gohr}}, \bibinfo {author} {\bibfnamefont {Marc}\ \bibnamefont {Strickert}},
  \ and\ \bibinfo {author} {\bibfnamefont {Ivo}\ \bibnamefont {Grosse}},\
  }\bibfield  {title} {\enquote {\bibinfo {title} {{Parsimonious higher-order
  hidden Markov models for improved array-CGH analysis with applications to
  Arabidopsis Thaliana}},}\ }\href {\doibase 10.1371/journal.pcbi.1002286}
  {\bibfield  {journal} {\bibinfo  {journal} {PLoS computational biology}\
  }\textbf {\bibinfo {volume} {8}},\ \bibinfo {pages} {e1002286} (\bibinfo
  {year} {2012})}\BibitemShut {NoStop}%
\bibitem [{\citenamefont {Singer}\ \emph {et~al.}(2014)\citenamefont {Singer},
  \citenamefont {Helic}, \citenamefont {Taraghi},\ and\ \citenamefont
  {Strohmaier}}]{sin14}%
  \BibitemOpen
  \bibfield  {author} {\bibinfo {author} {\bibfnamefont {Philipp}\ \bibnamefont
  {Singer}}, \bibinfo {author} {\bibfnamefont {Denis}\ \bibnamefont {Helic}},
  \bibinfo {author} {\bibfnamefont {Behnam}\ \bibnamefont {Taraghi}}, \ and\
  \bibinfo {author} {\bibfnamefont {Markus}\ \bibnamefont {Strohmaier}},\
  }\bibfield  {title} {\enquote {\bibinfo {title} {{Detecting memory and
  structure in human navigation patterns using Markov chain models of varying
  order}},}\ }\href {\doibase 10.1371/journal.pone.0114952} {\bibfield
  {journal} {\bibinfo  {journal} {PloS one}\ }\textbf {\bibinfo {volume} {9}},\
  \bibinfo {pages} {e102070} (\bibinfo {year} {2014})}\BibitemShut {NoStop}%
\bibitem [{\citenamefont {Meyer}\ and\ \citenamefont {Rieger}(2021)}]{mey21}%
  \BibitemOpen
  \bibfield  {author} {\bibinfo {author} {\bibfnamefont {Hugues}\ \bibnamefont
  {Meyer}}\ and\ \bibinfo {author} {\bibfnamefont {Heiko}\ \bibnamefont
  {Rieger}},\ }\bibfield  {title} {\enquote {\bibinfo {title} {Optimal
  non-{Markovian} search strategies with $n$-step memory},}\ }\href {\doibase
  10.1103/PhysRevLett.127.070601} {\bibfield  {journal} {\bibinfo  {journal}
  {Phys. Rev. Lett.}\ }\textbf {\bibinfo {volume} {127}},\ \bibinfo {pages}
  {070601} (\bibinfo {year} {2021})}\BibitemShut {NoStop}%
\bibitem [{\citenamefont {Strelioff}\ \emph {et~al.}(2007)\citenamefont
  {Strelioff}, \citenamefont {Crutchfield},\ and\ \citenamefont
  {H\"ubler}}]{crutchfield}%
  \BibitemOpen
  \bibfield  {author} {\bibinfo {author} {\bibfnamefont {Christopher~C.}\
  \bibnamefont {Strelioff}}, \bibinfo {author} {\bibfnamefont {James~P.}\
  \bibnamefont {Crutchfield}}, \ and\ \bibinfo {author} {\bibfnamefont
  {Alfred~W.}\ \bibnamefont {H\"ubler}},\ }\bibfield  {title} {\enquote
  {\bibinfo {title} {Inferring markov chains: Bayesian estimation, model
  comparison, entropy rate, and out-of-class modeling},}\ }\href {\doibase
  10.1103/PhysRevE.76.011106} {\bibfield  {journal} {\bibinfo  {journal} {Phys.
  Rev. E}\ }\textbf {\bibinfo {volume} {76}},\ \bibinfo {pages} {011106}
  (\bibinfo {year} {2007})}\BibitemShut {NoStop}%
\bibitem [{\citenamefont {Akaike}(1998)}]{aka98}%
  \BibitemOpen
  \bibfield  {author} {\bibinfo {author} {\bibfnamefont {Hirotogu}\
  \bibnamefont {Akaike}},\ }\bibfield  {title} {\enquote {\bibinfo {title}
  {Information theory and an extension of the maximum likelihood principle},}\
  }in\ \href@noop {} {\emph {\bibinfo {booktitle} {{Selected papers of hirotugu
  akaike}}}}\ (\bibinfo  {publisher} {Springer},\ \bibinfo {year} {1998})\ pp.\
  \bibinfo {pages} {199--213}\BibitemShut {NoStop}%
\bibitem [{\citenamefont {Tong}(1975)}]{ton75}%
  \BibitemOpen
  \bibfield  {author} {\bibinfo {author} {\bibfnamefont {Howell}\ \bibnamefont
  {Tong}},\ }\bibfield  {title} {\enquote {\bibinfo {title} {{Determination of
  the order of a Markov chain by Akaike's information criterion}},}\ }\href
  {\doibase 10.2307/3212863} {\bibfield  {journal} {\bibinfo  {journal}
  {Journal of applied probability}\ }\textbf {\bibinfo {volume} {12}},\
  \bibinfo {pages} {488--497} (\bibinfo {year} {1975})}\BibitemShut {NoStop}%
\bibitem [{\citenamefont {Katz}(1981)}]{kat81}%
  \BibitemOpen
  \bibfield  {author} {\bibinfo {author} {\bibfnamefont {Richard~W}\
  \bibnamefont {Katz}},\ }\bibfield  {title} {\enquote {\bibinfo {title} {{On
  some criteria for estimating the order of a Markov chain}},}\ }\href
  {\doibase doi.org/10.2307/1267787} {\bibfield  {journal} {\bibinfo  {journal}
  {Technometrics}\ }\textbf {\bibinfo {volume} {23}},\ \bibinfo {pages}
  {243--249} (\bibinfo {year} {1981})}\BibitemShut {NoStop}%
\bibitem [{\citenamefont {Weakliem}(1999)}]{wea99}%
  \BibitemOpen
  \bibfield  {author} {\bibinfo {author} {\bibfnamefont {David~L}\ \bibnamefont
  {Weakliem}},\ }\bibfield  {title} {\enquote {\bibinfo {title} {{A critique of
  the Bayesian information criterion for model selection}},}\ }\href {\doibase
  10.1177/0049124199027003002} {\bibfield  {journal} {\bibinfo  {journal}
  {Sociological Methods \& Research}\ }\textbf {\bibinfo {volume} {27}},\
  \bibinfo {pages} {359--397} (\bibinfo {year} {1999})}\BibitemShut {NoStop}%
\bibitem [{\citenamefont {Schmitt}\ and\ \citenamefont {Herzel}(1997)}]{sch97}%
  \BibitemOpen
  \bibfield  {author} {\bibinfo {author} {\bibfnamefont {Armin~O}\ \bibnamefont
  {Schmitt}}\ and\ \bibinfo {author} {\bibfnamefont {Hanspeter}\ \bibnamefont
  {Herzel}},\ }\bibfield  {title} {\enquote {\bibinfo {title} {Estimating the
  entropy of {DNA} sequences},}\ }\href {\doibase 10.1006/jtbi.1997.0493}
  {\bibfield  {journal} {\bibinfo  {journal} {Journal of theoretical biology}\
  }\textbf {\bibinfo {volume} {188}},\ \bibinfo {pages} {369--377} (\bibinfo
  {year} {1997})}\BibitemShut {NoStop}%
\bibitem [{\citenamefont {Chao}\ and\ \citenamefont {Shen}(2003)}]{chao}%
  \BibitemOpen
  \bibfield  {author} {\bibinfo {author} {\bibfnamefont {Anne}\ \bibnamefont
  {Chao}}\ and\ \bibinfo {author} {\bibfnamefont {Tsung-Jen}\ \bibnamefont
  {Shen}},\ }\bibfield  {title} {\enquote {\bibinfo {title} {{Nonparametric
  estimation of Shannon's diversity index when there are unseen species in
  sample}},}\ }\href {\doibase 10.1023/A:1026096204727} {\bibfield  {journal}
  {\bibinfo  {journal} {Environmental and Ecological Statistics}\ }\textbf
  {\bibinfo {volume} {10}},\ \bibinfo {pages} {429--443} (\bibinfo {year}
  {2003})}\BibitemShut {NoStop}%
\bibitem [{\citenamefont {H{\"a}ggstr{\"o}m}\ \emph {et~al.}(2002)\citenamefont
  {H{\"a}ggstr{\"o}m} \emph {et~al.}}]{hag02}%
  \BibitemOpen
  \bibfield  {author} {\bibinfo {author} {\bibfnamefont {Olle}\ \bibnamefont
  {H{\"a}ggstr{\"o}m}} \emph {et~al.},\ }\href@noop {} {\emph {\bibinfo {title}
  {{Finite Markov chains and algorithmic applications}}}},\ Vol.~\bibinfo
  {volume} {52}\ (\bibinfo  {publisher} {Cambridge University Press},\ \bibinfo
  {year} {2002})\BibitemShut {NoStop}%
\bibitem [{\citenamefont {Pompe}(1994)}]{pom94}%
  \BibitemOpen
  \bibfield  {author} {\bibinfo {author} {\bibfnamefont {B}~\bibnamefont
  {Pompe}},\ }\bibfield  {title} {\enquote {\bibinfo {title} {On some entropy
  methods in data analysis},}\ }\href {\doibase 10.1016/0960-0779(94)90019-1}
  {\bibfield  {journal} {\bibinfo  {journal} {Chaos, Solitons \& Fractals}\
  }\textbf {\bibinfo {volume} {4}},\ \bibinfo {pages} {83--96} (\bibinfo {year}
  {1994})}\BibitemShut {NoStop}%
\bibitem [{\citenamefont {Shannon}(1948)}]{shannon}%
  \BibitemOpen
  \bibfield  {author} {\bibinfo {author} {\bibfnamefont {C.~E.}\ \bibnamefont
  {Shannon}},\ }\bibfield  {title} {\enquote {\bibinfo {title} {A mathematical
  theory of communication},}\ }\href {\doibase
  10.1002/j.1538-7305.1948.tb01338.x} {\bibfield  {journal} {\bibinfo
  {journal} {Bell System Technical Journal}\ }\textbf {\bibinfo {volume}
  {27}},\ \bibinfo {pages} {379--423} (\bibinfo {year} {1948})}\BibitemShut
  {NoStop}%
\bibitem [{\citenamefont {Park}\ and\ \citenamefont {Pande}(2006)}]{par06}%
  \BibitemOpen
  \bibfield  {author} {\bibinfo {author} {\bibfnamefont {Sanghyun}\
  \bibnamefont {Park}}\ and\ \bibinfo {author} {\bibfnamefont {Vijay~S}\
  \bibnamefont {Pande}},\ }\bibfield  {title} {\enquote {\bibinfo {title}
  {{Validation of Markov state models using Shannon's entropy}},}\ }\href
  {\doibase 10.1063/1.2166393} {\bibfield  {journal} {\bibinfo  {journal} {The
  Journal of chemical physics}\ }\textbf {\bibinfo {volume} {124}},\ \bibinfo
  {pages} {054118} (\bibinfo {year} {2006})}\BibitemShut {NoStop}%
\bibitem [{\citenamefont {Cover}\ and\ \citenamefont {Thomas}(2006)}]{cover}%
  \BibitemOpen
  \bibfield  {author} {\bibinfo {author} {\bibfnamefont {T.M.}\ \bibnamefont
  {Cover}}\ and\ \bibinfo {author} {\bibfnamefont {J.A.}\ \bibnamefont
  {Thomas}},\ }\href@noop {} {\emph {\bibinfo {title} {{Elements of Information
  Theory}}}}\ (\bibinfo  {publisher} {John Wiley and Sons},\ \bibinfo {year}
  {2006})\BibitemShut {NoStop}%
\bibitem [{\citenamefont {Sch{\"u}rmann}(2004)}]{Sch_rmann_2004}%
  \BibitemOpen
  \bibfield  {author} {\bibinfo {author} {\bibfnamefont {Thomas}\ \bibnamefont
  {Sch{\"u}rmann}},\ }\bibfield  {title} {\enquote {\bibinfo {title} {Bias
  analysis in entropy estimation},}\ }\href {\doibase
  10.1088/0305-4470/37/27/l02} {\bibfield  {journal} {\bibinfo  {journal}
  {Journal of Physics A: Mathematical and General}\ }\textbf {\bibinfo {volume}
  {37}},\ \bibinfo {pages} {L295--L301} (\bibinfo {year} {2004})}\BibitemShut
  {NoStop}%
\bibitem [{\citenamefont {Paninski}(2003)}]{paninski}%
  \BibitemOpen
  \bibfield  {author} {\bibinfo {author} {\bibfnamefont {Liam}\ \bibnamefont
  {Paninski}},\ }\bibfield  {title} {\enquote {\bibinfo {title} {Estimation of
  entropy and mutual information},}\ }\href {\doibase
  10.1162/089976603321780272} {\bibfield  {journal} {\bibinfo  {journal}
  {Neural Computation}\ }\textbf {\bibinfo {volume} {15}},\ \bibinfo {pages}
  {1191--1253} (\bibinfo {year} {2003})}\BibitemShut {NoStop}%
\bibitem [{\citenamefont {Contreras~Rodríguez}\ \emph
  {et~al.}(2021)\citenamefont {Contreras~Rodríguez}, \citenamefont
  {Madarro-Capó}, \citenamefont {Legón-Pérez}, \citenamefont {Rojas},\ and\
  \citenamefont {Sosa-Gómez}}]{contreras}%
  \BibitemOpen
  \bibfield  {author} {\bibinfo {author} {\bibfnamefont {Lianet}\ \bibnamefont
  {Contreras~Rodríguez}}, \bibinfo {author} {\bibfnamefont {Evaristo~José}\
  \bibnamefont {Madarro-Capó}}, \bibinfo {author} {\bibfnamefont
  {Carlos~Miguel}\ \bibnamefont {Legón-Pérez}}, \bibinfo {author}
  {\bibfnamefont {Omar}\ \bibnamefont {Rojas}}, \ and\ \bibinfo {author}
  {\bibfnamefont {Guillermo}\ \bibnamefont {Sosa-Gómez}},\ }\bibfield  {title}
  {\enquote {\bibinfo {title} {Selecting an effective entropy estimator for
  short sequences of bits and bytes with maximum entropy},}\ }\href {\doibase
  10.3390/e23050561} {\bibfield  {journal} {\bibinfo  {journal} {Entropy}\
  }\textbf {\bibinfo {volume} {23}} (\bibinfo {year} {2021}),\
  10.3390/e23050561}\BibitemShut {NoStop}%
\bibitem [{\citenamefont {Horvitz}\ and\ \citenamefont
  {Thompson}(1952)}]{horvitz}%
  \BibitemOpen
  \bibfield  {author} {\bibinfo {author} {\bibfnamefont {D.~G.}\ \bibnamefont
  {Horvitz}}\ and\ \bibinfo {author} {\bibfnamefont {D.~J.}\ \bibnamefont
  {Thompson}},\ }\bibfield  {title} {\enquote {\bibinfo {title} {A
  generalization of sampling without replacement from a finite universe},}\
  }\href {\doibase 10.1080/01621459.1952.10483446} {\bibfield  {journal}
  {\bibinfo  {journal} {Journal of the American Statistical Association}\
  }\textbf {\bibinfo {volume} {47}},\ \bibinfo {pages} {663--685} (\bibinfo
  {year} {1952})}\BibitemShut {NoStop}%
\bibitem [{\citenamefont {Good}(1953)}]{good}%
  \BibitemOpen
  \bibfield  {author} {\bibinfo {author} {\bibfnamefont {I.J.}\ \bibnamefont
  {Good}},\ }\bibfield  {title} {\enquote {\bibinfo {title} {The population
  frequencies of species and the estimation of population parameters},}\ }\href
  {\doibase 10.2307/2333344} {\bibfield  {journal} {\bibinfo  {journal}
  {Biometrika}\ }\textbf {\bibinfo {volume} {40}},\ \bibinfo {pages} {237--64}
  (\bibinfo {year} {1953})}\BibitemShut {NoStop}%
\bibitem [{\citenamefont {Wan}\ \emph {et~al.}(2005)\citenamefont {Wan},
  \citenamefont {Zhang},\ and\ \citenamefont {Barrow}}]{doi:10.3137/ao.430102}%
  \BibitemOpen
  \bibfield  {author} {\bibinfo {author} {\bibfnamefont {Hui}\ \bibnamefont
  {Wan}}, \bibinfo {author} {\bibfnamefont {Xuebin}\ \bibnamefont {Zhang}}, \
  and\ \bibinfo {author} {\bibfnamefont {Elaine~M.}\ \bibnamefont {Barrow}},\
  }\bibfield  {title} {\enquote {\bibinfo {title} {{Stochastic modelling of
  daily precipitation for Canada}},}\ }\href {\doibase 10.3137/ao.430102}
  {\bibfield  {journal} {\bibinfo  {journal} {Atmosphere-Ocean}\ }\textbf
  {\bibinfo {volume} {43}},\ \bibinfo {pages} {23--32} (\bibinfo {year}
  {2005})}\BibitemShut {NoStop}%
\bibitem [{\citenamefont {{Wilson Kemsley}}\ \emph {et~al.}(2021)\citenamefont
  {{Wilson Kemsley}}, \citenamefont {Osborn}, \citenamefont {Dorling},
  \citenamefont {Wallace},\ and\ \citenamefont {Parker}}]{WilsonKemsley2021}%
  \BibitemOpen
  \bibfield  {author} {\bibinfo {author} {\bibfnamefont {Sarah}\ \bibnamefont
  {{Wilson Kemsley}}}, \bibinfo {author} {\bibfnamefont {Timothy~J.}\
  \bibnamefont {Osborn}}, \bibinfo {author} {\bibfnamefont {Stephen~R.}\
  \bibnamefont {Dorling}}, \bibinfo {author} {\bibfnamefont {Craig}\
  \bibnamefont {Wallace}}, \ and\ \bibinfo {author} {\bibfnamefont {Joanne}\
  \bibnamefont {Parker}},\ }\bibfield  {title} {\enquote {\bibinfo {title}
  {{Selecting {M}arkov chain orders for generating daily precipitation series
  across different K{\"{o}}ppen climate regimes}},}\ }\href {\doibase
  10.1002/joc.7175} {\bibfield  {journal} {\bibinfo  {journal} {International
  Journal of Climatology}\ }\textbf {\bibinfo {volume} {41}},\ \bibinfo {pages}
  {6223--6237} (\bibinfo {year} {2021})}\BibitemShut {NoStop}%
\bibitem [{\citenamefont {Schwarz}(1978)}]{bic}%
  \BibitemOpen
  \bibfield  {author} {\bibinfo {author} {\bibfnamefont {Gideon}\ \bibnamefont
  {Schwarz}},\ }\bibfield  {title} {\enquote {\bibinfo {title} {Estimating the
  dimension of a model},}\ }\href {\doibase 10.1214/aos/1176344136} {\bibfield
  {journal} {\bibinfo  {journal} {The Annals of Statistics}\ }\textbf {\bibinfo
  {volume} {6}},\ \bibinfo {pages} {461--464} (\bibinfo {year}
  {1978})}\BibitemShut {NoStop}%
\bibitem [{\citenamefont {Menne}\ \emph {et~al.}(2012)\citenamefont {Menne},
  \citenamefont {Durre}, \citenamefont {Vose}, \citenamefont {Gleason},\ and\
  \citenamefont {Houston}}]{men12}%
  \BibitemOpen
  \bibfield  {author} {\bibinfo {author} {\bibfnamefont {Matthew~J}\
  \bibnamefont {Menne}}, \bibinfo {author} {\bibfnamefont {Imke}\ \bibnamefont
  {Durre}}, \bibinfo {author} {\bibfnamefont {Russell~S}\ \bibnamefont {Vose}},
  \bibinfo {author} {\bibfnamefont {Byron~E}\ \bibnamefont {Gleason}}, \ and\
  \bibinfo {author} {\bibfnamefont {Tamara~G}\ \bibnamefont {Houston}},\
  }\bibfield  {title} {\enquote {\bibinfo {title} {An overview of the global
  historical climatology network-daily database},}\ }\href {\doibase
  10.1175/JTECH-D-11-00103.1} {\bibfield  {journal} {\bibinfo  {journal}
  {Journal of atmospheric and oceanic technology}\ }\textbf {\bibinfo {volume}
  {29}},\ \bibinfo {pages} {897--910} (\bibinfo {year} {2012})}\BibitemShut
  {NoStop}%
\bibitem [{\citenamefont {Nemenman}\ \emph {et~al.}(2001)\citenamefont
  {Nemenman}, \citenamefont {Shafee},\ and\ \citenamefont {Bialek}}]{nsb}%
  \BibitemOpen
  \bibfield  {author} {\bibinfo {author} {\bibfnamefont {Ilya}\ \bibnamefont
  {Nemenman}}, \bibinfo {author} {\bibfnamefont {F.}~\bibnamefont {Shafee}}, \
  and\ \bibinfo {author} {\bibfnamefont {William}\ \bibnamefont {Bialek}},\
  }\bibfield  {title} {\enquote {\bibinfo {title} {Entropy and inference,
  revisited},}\ }in\ \href@noop {} {\emph {\bibinfo {booktitle} {Advances in
  Neural Information Processing Systems}}},\ Vol.~\bibinfo {volume} {14},\
  \bibinfo {editor} {edited by\ \bibinfo {editor} {\bibfnamefont
  {T.}~\bibnamefont {Dietterich}}, \bibinfo {editor} {\bibfnamefont
  {S.}~\bibnamefont {Becker}}, \ and\ \bibinfo {editor} {\bibfnamefont
  {Z.}~\bibnamefont {Ghahramani}}}\ (\bibinfo  {publisher} {MIT Press},\
  \bibinfo {year} {2001})\BibitemShut {NoStop}%
\bibitem [{\citenamefont {Archer}\ \emph {et~al.}(2013)\citenamefont {Archer},
  \citenamefont {Park},\ and\ \citenamefont {Pillow}}]{archer}%
  \BibitemOpen
  \bibfield  {author} {\bibinfo {author} {\bibfnamefont {Evan~W}\ \bibnamefont
  {Archer}}, \bibinfo {author} {\bibfnamefont {Il~Memming}\ \bibnamefont
  {Park}}, \ and\ \bibinfo {author} {\bibfnamefont {Jonathan~W}\ \bibnamefont
  {Pillow}},\ }\bibfield  {title} {\enquote {\bibinfo {title} {Bayesian entropy
  estimation for binary spike train data using parametric prior knowledge},}\
  }in\ \href@noop {} {\emph {\bibinfo {booktitle} {Advances in Neural
  Information Processing Systems}}},\ Vol.~\bibinfo {volume} {26},\ \bibinfo
  {editor} {edited by\ \bibinfo {editor} {\bibfnamefont {C.J.}\ \bibnamefont
  {Burges}}, \bibinfo {editor} {\bibfnamefont {L.}~\bibnamefont {Bottou}},
  \bibinfo {editor} {\bibfnamefont {M.}~\bibnamefont {Welling}}, \bibinfo
  {editor} {\bibfnamefont {Z.}~\bibnamefont {Ghahramani}}, \ and\ \bibinfo
  {editor} {\bibfnamefont {K.Q.}\ \bibnamefont {Weinberger}}}\ (\bibinfo
  {publisher} {Curran Associates, Inc.},\ \bibinfo {year} {2013})\BibitemShut
  {NoStop}%
\bibitem [{\citenamefont {Hausser}\ and\ \citenamefont {Strimmer}(2009)}]{shr}%
  \BibitemOpen
  \bibfield  {author} {\bibinfo {author} {\bibfnamefont {Jean}\ \bibnamefont
  {Hausser}}\ and\ \bibinfo {author} {\bibfnamefont {Korbinian}\ \bibnamefont
  {Strimmer}},\ }\bibfield  {title} {\enquote {\bibinfo {title} {Entropy
  inference and the {J}ames-{S}tein estimator, with application to nonlinear
  gene association networks},}\ }\href {\doibase 10.5555/1577069.1755833}
  {\bibfield  {journal} {\bibinfo  {journal} {Journal of Machine Learning
  Research}\ }\textbf {\bibinfo {volume} {10}},\ \bibinfo {pages} {1469--1484}
  (\bibinfo {year} {2009})}\BibitemShut {NoStop}%
\bibitem [{\citenamefont {Grassberger}(2022)}]{grassberger}%
  \BibitemOpen
  \bibfield  {author} {\bibinfo {author} {\bibfnamefont {Peter}\ \bibnamefont
  {Grassberger}},\ }\bibfield  {title} {\enquote {\bibinfo {title} {{On
  Generalized Sch\"urmann Entropy Estimators}},}\ }\href {\doibase
  10.3390/e24050680} {\bibfield  {journal} {\bibinfo  {journal} {Entropy}\
  }\textbf {\bibinfo {volume} {24}} (\bibinfo {year} {2022}),\
  10.3390/e24050680}\BibitemShut {NoStop}%
\end{thebibliography}%


\onecolumngrid

\appendix
\setcounter{figure}{0}
\setcounter{equation}{0}
\renewcommand{\thefigure}{A\arabic{figure}} 		

\section{}
\label{sec:appA}
\subsection{Proof that if a sequence has memory $m$, then $H_n$ is a linear function for $n \geq m$ \label{appendix}}
\label{subsec:appA1}
Since we consider only homogeneous sequences we will omit the time variable in this appendix. We will use the notation $p(x_1,\ldots,x_s)\equiv P(X_1=x_1,\ldots,X_s=x_s)$, with $x_i\in \lbrace z_j \rbrace_{1\leq j\leq L}$.

Let us calculate the difference between the block entropies $H_{n+1}-H_n$. Using $p(x_1,\ldots,x_{n+1})=p(x_{n+1}|x_1,\ldots,x_{n}) p(x_1,\ldots,x_{n})$, we obtain
\begin{align}
\begin{split}
H_{n+1}-H_n &= -\sum_{x_1,\ldots,x_{n+1}}p(x_1,\ldots,x_{n+1})\log(p(x_1,\ldots,x_{n+1})) + \sum_{x_1,\ldots,x_{n}}p(x_1,\ldots,x_{n})\log(p(x_1,\ldots,x_{n})) \\
&= -\sum_{x_1,\ldots,x_{n+1}}p(x_1,\ldots,x_{n+1})\log(p(x_{n+1}|x_1,\ldots,x_n)) - \sum_{x_1,\ldots,x_{n+1}}p(x_1,\ldots,x_{n+1})\log(p(x_1,\ldots,x_{n})) + \\ 
&+ \sum_{x_1,\ldots,x_{n}}p(x_1,\ldots,x_{n})\log(p(x_1,\ldots,x_{n})).
\end{split}
\label{eq:a1}
\end{align} 
Because $\sum\limits_{x_{n+1}}p(x_1,\ldots,x_{n+1}) = p(x_1,\ldots,x_{n})$, the last two terms of the Eq.~\eqref{eq:a1} cancel out. Thus,
\begin{align}
\begin{split}
H_{n+1}-H_n &= -\sum_{x_1,\ldots,x_{n+1}}p(x_1,\ldots,x_{n+1})\log(p(x_{n+1}|x_1,\ldots,x_n)).
\end{split}
\label{eq:a2}
\end{align}
This is a general result, valid for any kind of sequence.
Now, for a sequence of memory $m$ and for $n \geq m$, one has $p(x_{n+1}|x_1,\ldots,x_n) = p(x_{n+1}|x_{n-m+1},\ldots,x_n)$. Then, Eq.~\eqref{eq:a2} becomes
\begin{align}
\begin{split}
H_{n+1}-H_n &= -\sum_{x_1,\ldots,x_{n+1}}p(x_1,\ldots,x_{n+1})\log(p(x_{n+1}|x_{n-m+1},\ldots,x_n)) \\
&= -\sum_{x_{n-m+1},...,x_{n+1}}p(x_{n-m+1},...,x_{n+1})\log(p(x_{n+1}|x_{n-m+1},...,x_{n})). 
\end{split}
\label{eq:a3}
\end{align}
For homogeneous sequences, we can shift all indices on the right-hand side of Eq.~\eqref{eq:a3} by an amount $n-m\ge 0$:
\begin{align}
\begin{split}
H_{n+1}-H_n &= -\sum_{x_{1},...,x_{m+1}}p(x_{1},...,x_{m+1})\log(p(x_{m+1}|x_{1},...,x_{m})). 
\end{split}
\label{eq:a4}
\end{align}
As far as $n\ge m$, the right-hand side of Eq.~\eqref{eq:a4} is independent of $n$. Making the replacement $n\rightarrow m$ we arrive at
\begin{align}
H_{n+1}-H_n &= H_{m+1}-H_m, \quad n \geq m,
\label{eq:a5}
\end{align}
which proves that the dependence of $H_n$ on $n$ is linear for $n \geq m$, i.e., $H_n=an+b$ with constant parameters $a$ and $b$. 

\subsection{Proof that if $H_n$ is linear for $n \geq m$, then the sequence has memory $m$}

Let $H_n$ be linear for $n \geq m$. Then, we can write $H_n=an+b$ or
\begin{equation}
H_{n+1}-H_n = H_{m+1}-H_m, \quad n \geq m. 
\label{eq:b1}
\end{equation}
Using the general result of Eq.~\eqref{eq:a2} on the right-hand side of Eq.~\eqref{eq:b1},
we find
\begin{align}
H_{n+1}-H_n = -\sum_{x_{1},...,x_{m+1}}p(x_{1},...,x_{m+1})\log(p(x_{m+1}|x_{1},...,x_{m})).
\label{eq:b2}
\end{align}
For homogeneous sequences, we can shift all indices on the right-hand side of Eq.~\eqref{eq:b2} by an amount $n-m\ge 0$:
\begin{align}
\begin{split}
H_{n+1}-H_n &= -\sum_{x_{n-m+1},...,x_{n+1}}p(x_{n-m+1},...,x_{n+1})\log(p(x_{n+1}|x_{n-m+1},...,x_{n})) \\
&= -\sum_{x_{1},...,x_{n+1}}p(x_{1},...,x_{n+1})\log(p(x_{n+1}|x_{n-m+1},...,x_{n})).
\end{split}
\label{eq:b3}
\end{align}
We now apply Eq.~\eqref{eq:a2} on the left-hand side of Eq.~\eqref{eq:b3}:
\begin{align}
\sum_{x_{1},...,x_{n+1}}p(x_{1},...,x_{n+1})\log(p(x_{n+1}|x_{1},...,x_{n}))= \sum_{x_{1},...,x_{n+1}}p(x_{1},...,x_{n+1})\log(p(x_{n+1}|x_{n-m+1},...,x_{n})),
\label{eq:b4}
\end{align}
which implies that
\begin{align}
\sum_{x_{1},...,x_{n+1}}p(x_{1},...,x_{n+1})\log \left(\dfrac{p(x_{n+1}|x_{1},...,x_{n})}{p(x_{n+1}|x_{n-m+1},...,x_{n})}\right) = 0,
\label{eq:b5}
\end{align}
or
\begin{align}
\sum_{x_{1},...,x_{n}}p(x_{1},...,x_{n})\sum_{x_{n+1}}p(x_{n+1}|x_{1},...,x_{n})\log \left(\dfrac{p(x_{n+1}|x_{1},...,x_{n})}{p(x_{n+1}|x_{n-m+1},...,x_{n})}\right) = 0.
\label{eq:b6}
\end{align}
Because of the log sum inequality~\cite{cover}, we know that the second sum of Eq.~\eqref{eq:b6} is $\geq 0$ and generally $p(x_{1},...,x_{n}) > 0$. Hence, Eq.~\eqref{eq:b6} holds only if 
\begin{align}
\sum_{x_{n+1}}p(x_{n+1}|x_{1},...,x_{n})\log \left(\dfrac{p(x_{n+1}|x_{1},...,x_{n})}{p(x_{n+1}|x_{n-m+1},...,x_{n})}\right) = 0 \quad \forall x_1,\ldots,x_n.
\label{eq:b7}
\end{align}
Further, due to the log sum inequality, Eq.~\eqref{eq:b7} is valid provided that $p(x_{n+1}|x_{1},...,x_{n})=p(x_{n+1}|x_{n-m+1},...,x_{n}) \quad \forall \: x_{n+1},x_{1},...,x_{n}, \quad \forall n \geq m$, which means that the sequence has memory $m$.

\renewcommand{\thefigure}{B\arabic{figure}} 		

\section{}
\label{sec:appB}

Given a sequence $\MS$ of length $N$, we calculate the probability that we observe the block $b_i^{(n)}$ of size $n$ as follows,
\begin{equation}
P(b_i^{(n)}\in \MS) = 1-(1-p(b_i^{(n)}))^{N_n}.
\label{eq:inc_prob2}
\end{equation}
Even though Eq.~\eqref{eq:inc_prob2} is exact only when $n=1$ and the sequence memory is $m=0$, we have checked in all our simulations that the corrections introduced by correlations when $m>0$ and $n>1$ can be neglected if $N\gg n$.

\begin{figure}[t]
\scalebox{.5}{
\subfloat{\includegraphics[width=0.48\textwidth]{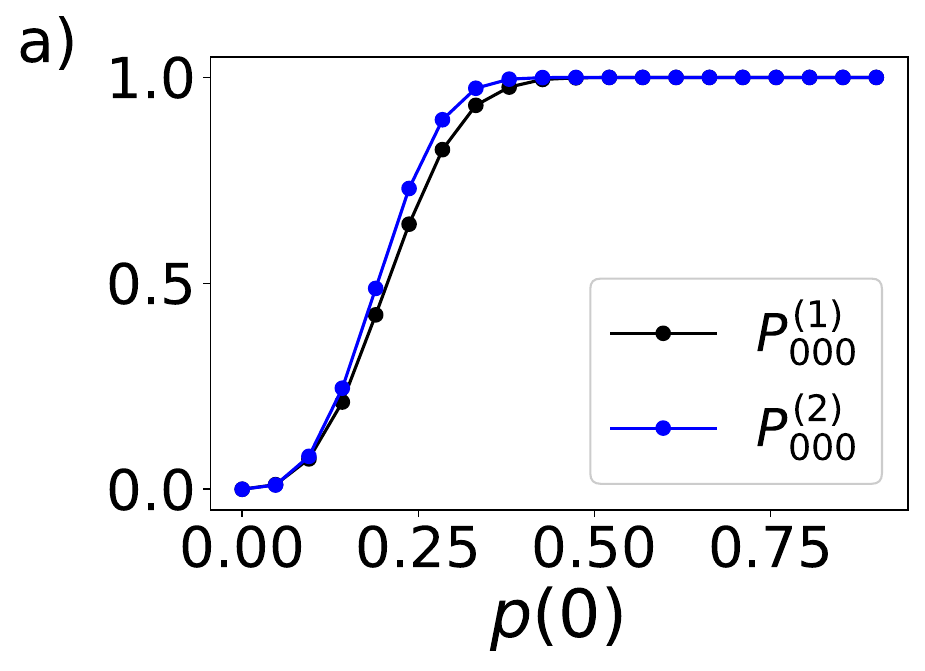}}
\hfill
\subfloat{\includegraphics[width=0.48\textwidth]{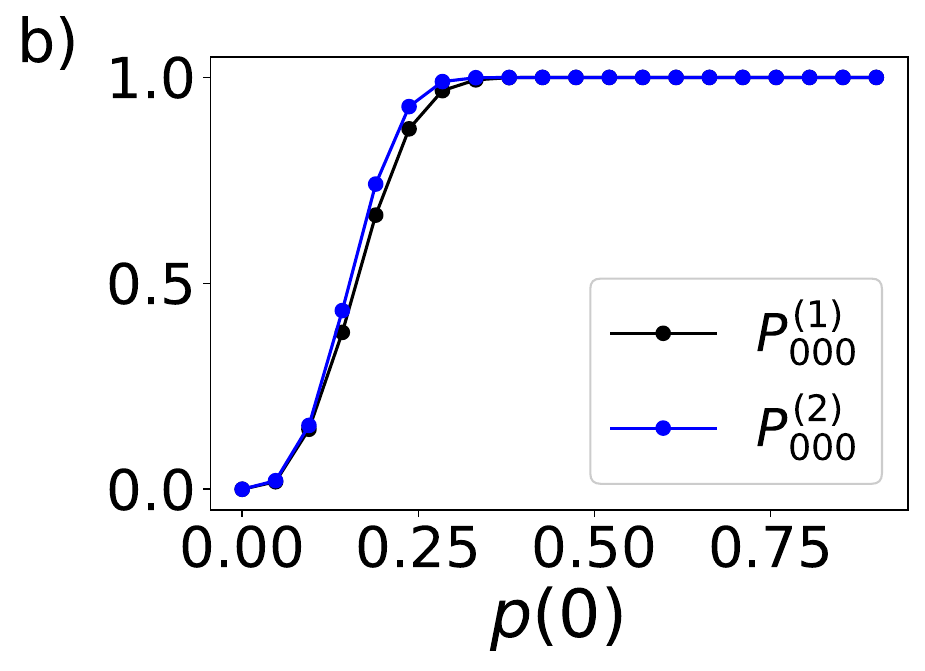}}
\newline
 \subfloat{\includegraphics[width=0.48\textwidth]{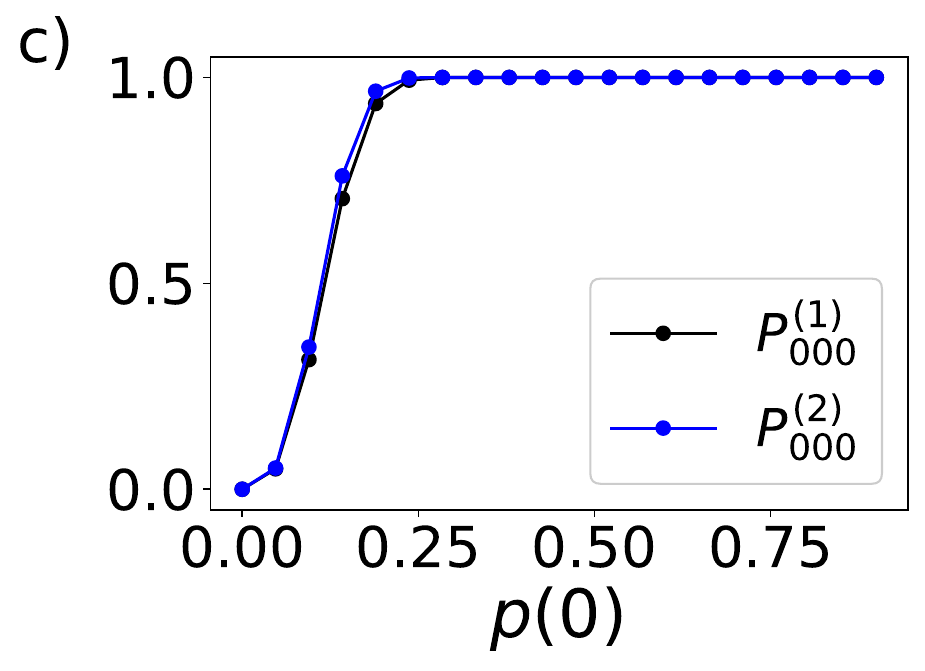}}
\hfill
 \subfloat{\includegraphics[width=0.48\textwidth]{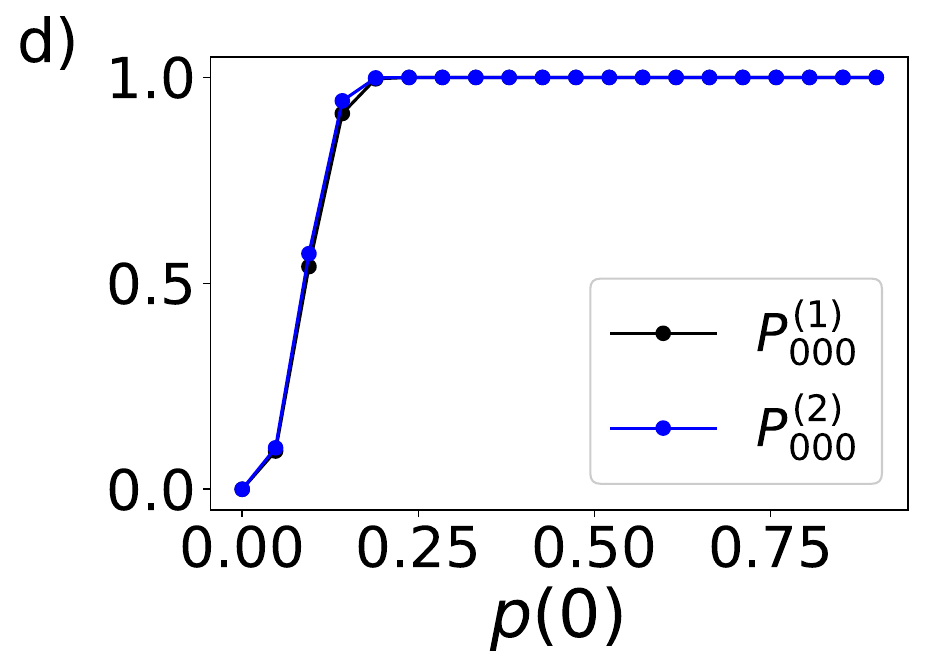}}
 }
 \caption{$P_{000}^{(1)}$ (black lines) for binary sequences with memory $m=0$ as a function of the probability $p(0)$ for occurrence of outcome $0$. For each combination of $N$ and $p(0)$, we generate $K=10^4$ numerical sequences and calculate $P_{000}^{(1)}$ from $K_{000}/K$, where $K_{000}$ is the number of sequences where the block $(0,0,0)$ appears at least once. We also plot with blue lines $P_{000}^{(2)}=1-(1-(p(0))^3)^{N-2}$. We show results for $N=100$ in (a), $200$ in (b), $500$ in (c) and $1000$ in (d). We observe that the black and blue curves overlap as $N$ grows.}
 \label{fig:inc_prob_m=0}
\end{figure}

\begin{figure}[b]
\scalebox{.5}{
\subfloat{\includegraphics[width=0.48\textwidth]{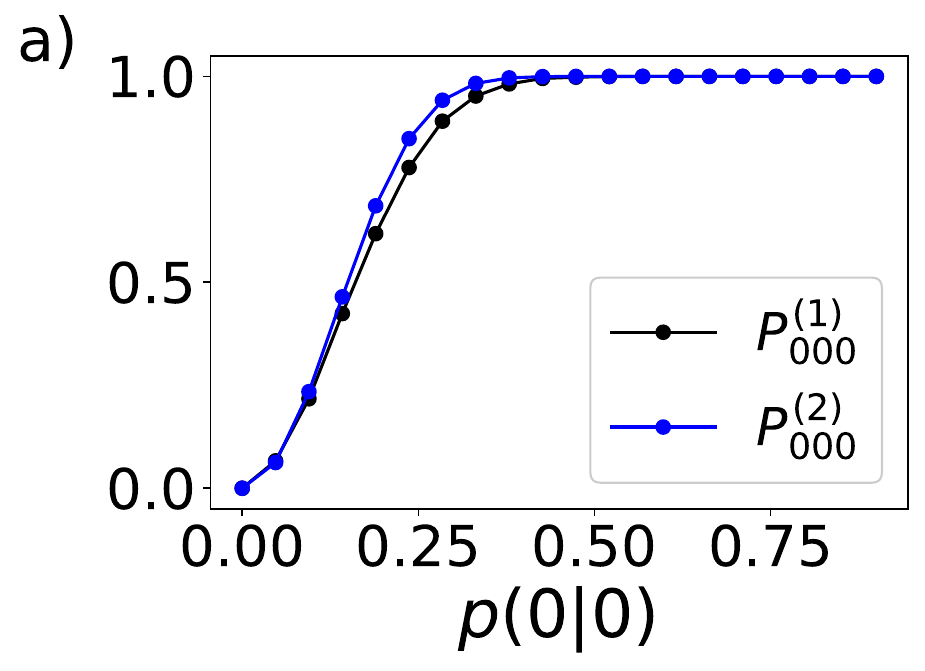}}
\hfill
\subfloat{\includegraphics[width=0.48\textwidth]{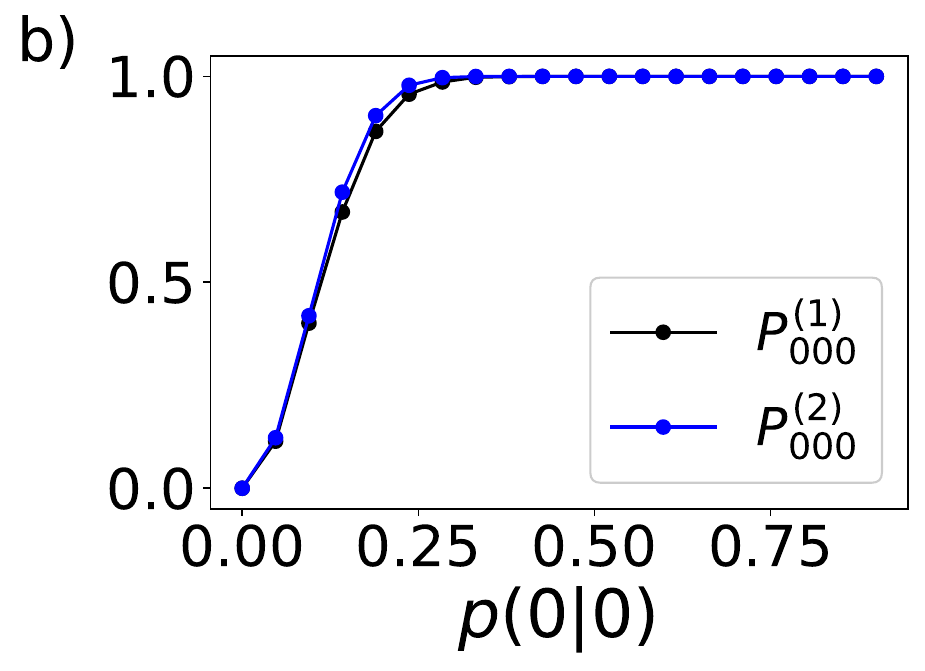}}
\newline
 \subfloat{\includegraphics[width=0.48\textwidth]{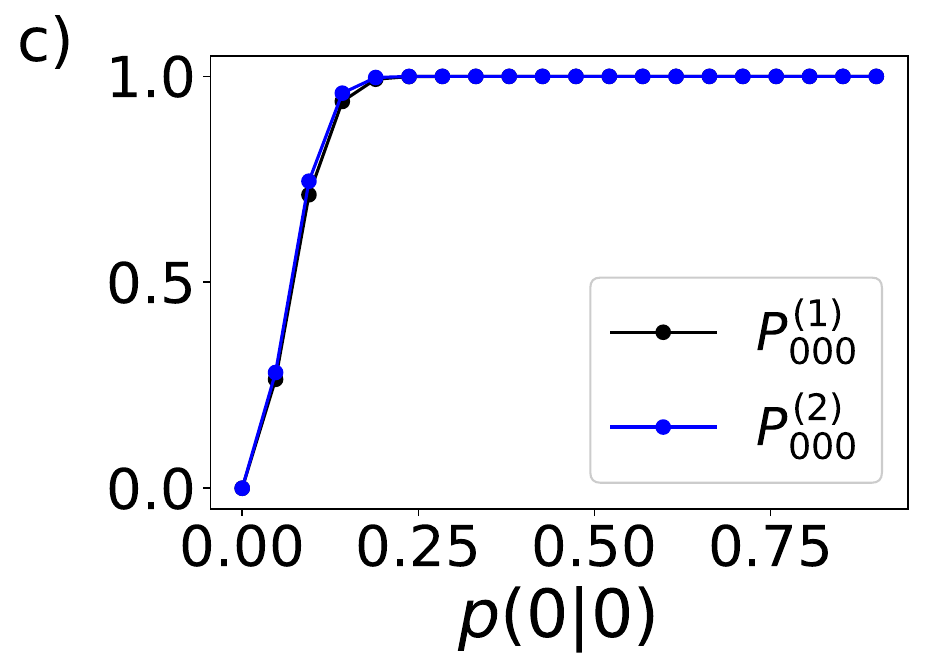}}
\hfill
 \subfloat{\includegraphics[width=0.48\textwidth]{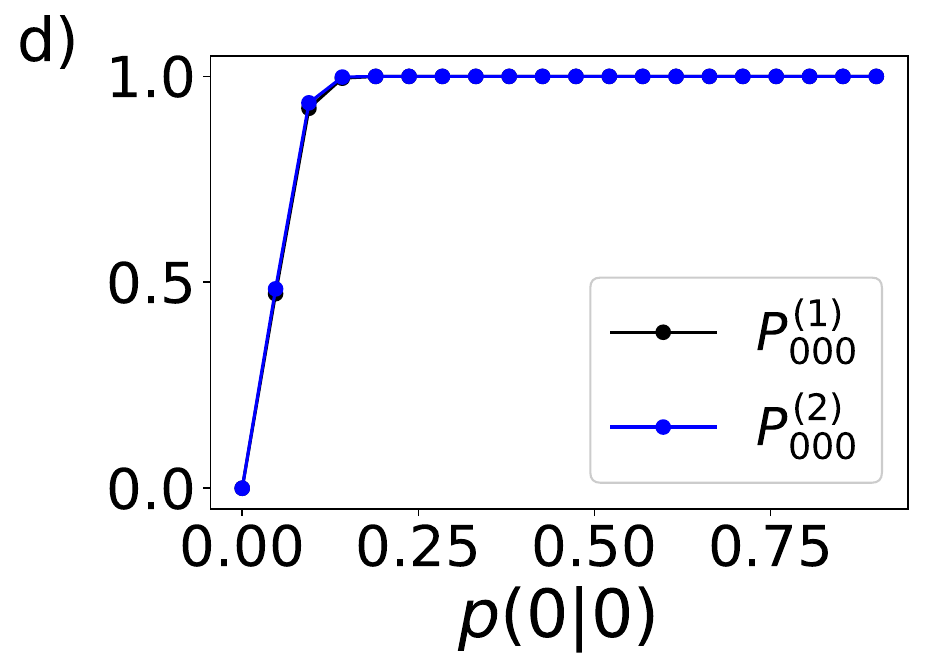}}
 }
 \caption{$P_{000}^{(1)}$ (black lines) for binary sequences with memory $m=1$
 as a function of the conditional probability $p(0|0)$. For each combination of $N$ and $p(0|0)$, we generate $K=10^4$ numerical sequences and calculate $P_{000}^{(1)}$ from  $K_{000}/K$, where $K_{000}$ is the number of sequences where the block $(0,0,0)$ appears at least once. We also plot with blue lines $P_{000}^{(2)}$ calculated from the transition probabilities. For all the cases considered we tale $p(1|1)=0.6$. We show results for $N=100$ in (a), $200$ in (b), $500$ in (c) and $1000$ in (d). We observe that the black and blue curves overlap as $N$ grows.} 
 \label{fig:inc_prob_m=1}
\end{figure}

As an example, we now assess the probability that the block $(X_s=0,X_{s+1}=0,X_{s+2}=0)\equiv (0,0,0)$ appears in $\MS$ by generating $K$ numerical sequences for fixed parameters $L$, $m$ and $N$, and a particular set of transition probabilities.
If the block $(0,0,0)$ appears in $K_{000}$ of those sequences, then $P_{000}^{(1)}\equiv \hat{P}((0,0,0)\in \MS) = K_{000}/K$.
We note that for $K\gg 1$ $P_{000}^{(1)}\simeq P((0,0,0)\in \MS)$.
We compare this result with the value $P_{000}^{(2)}\equiv 1-(1-p(0,0,0))^{N-2}$, where $p(0,0,0)$ is the probability of occurrence of the block $(0,0,0)$, which is computed from the transition probabilities.

In Fig.~\ref{fig:inc_prob_m=0} we plot the results for both $P_{000}^{(1)}$ and $P_{000}^{(2)}$ obtained from for binary sequences with $m=0$, as a function of $p(0)$. We consider the cases $N=100,200,500,1000$ in Figs.~\ref{fig:inc_prob_m=0}(a), (b), (c) and (d), respectively,
for $K=10^4$ repetitions.

In Fig.~\ref{fig:inc_prob_m=1} we show similar curves but now considering sequences with memory $m=1$. For each case, we fix $p(1|1)=0.6$ and vary $p(0|0)$ between $0$ and $0.9$.

The results plotted in both Figs.~\ref{fig:inc_prob_m=0} and \ref{fig:inc_prob_m=1} clearly show that as $N$ increases the difference between $P_{000}^{(1)}$ and $P_{000}^{(2)}$ vanishes.  
We have verified with our simulations that this holds for every block sequence and for different values of $m$. Therefore, Eq.~\eqref{eq:inc_prob2} is an excellent approximation
when the size of the sequence is much larger than the size of the block, regardless
of the memory value.

\setcounter{figure}{0}
\renewcommand{\thefigure}{C\arabic{figure}} 		

\section{}
\label{sec:appC}

In Fig.~\ref{fig:m=2,3} we show the exact entropy $H_n$ for binary systems with memory $m=2$ in (a) and $m=3$ in (b). In both cases the transition probabilities are chosen randomly from an uniform distribution $(0,1)$. As a comparison, we also show the results obtained with the MLE estimator given by Eq.~\ref{eq:HMLE} (green line and dots), the Chao-Shen estimator given by Eq.~\ref{eq:HCS} (blue line and dots) and our coverage-adjusted estimator proposed in Eq.~\ref{eq:Hnew} (red line and dots), all calculated from a sequence of $N=10^4$ data points generated numerically.

\begin{figure}[!htb]
\centering
{\includegraphics[width=.45\textwidth]{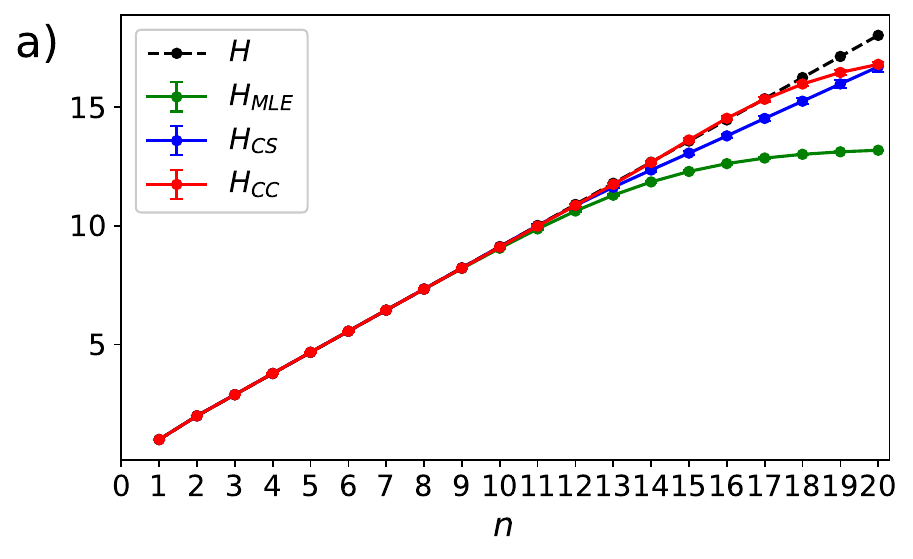}}\quad
{\includegraphics[width=.45\textwidth]{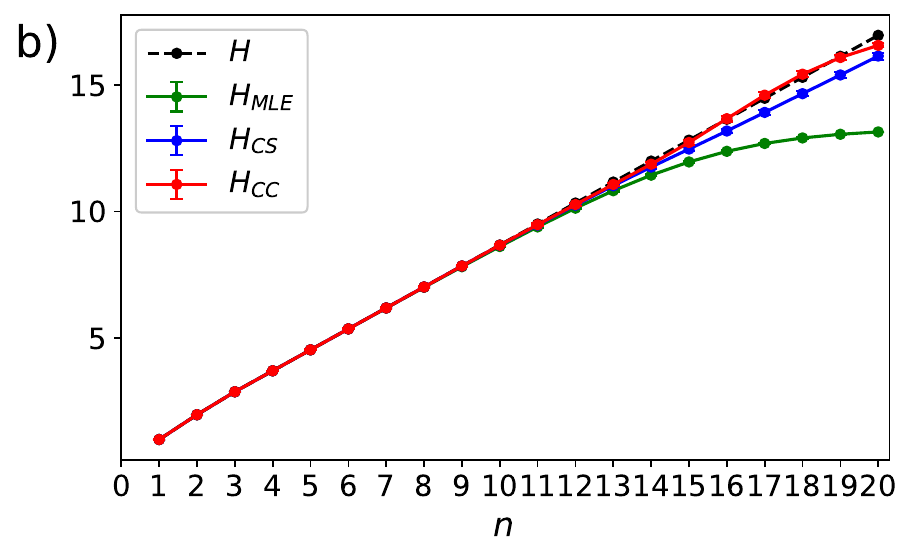}}
\caption{Exact Shannon entropy per block of size $n$ (dotted line) for binary systems with $m=2$ in (a) and $m=3$ in (b) with fixed transition probabilities chosen randomly from an uniform distribution. A sequence of $N=10^4$ realizations is numerically generated from which we calculate three different entropy estimators discussed in the main text. The best performance is shown for the estimator that takes into account correlations (red dots).}
\label{fig:m=2,3}
\end{figure}

\end{document}